\documentclass[letterpaper,twocolumn,10pt]{article}
\usepackage{usenix2019_v3}

\usepackage{tikz}
\usepackage{amsmath}

\usepackage{filecontents}
\usepackage{times}
\usepackage{amsmath}
\usepackage{amssymb}
\usepackage{enumitem}
\usepackage{latexsym}
\usepackage{subcaption}
\usepackage{xcolor}
\usepackage{tabularray}
\usepackage{multirow}
\usepackage{booktabs}
\usepackage[most]{tcolorbox}

\definecolor{FindingBlue}{HTML}{1E3A5F}

\newtcolorbox{findingbox}{
  colback=FindingBlue!2,
  colframe=FindingBlue,
  boxrule=1pt,
  arc=1.5pt,
  boxsep=0pt,
  left=1.5pt,
  right=1.5pt,
}

\usepackage[T1]{fontenc}

\usepackage[utf8]{inputenc}

\usepackage{microtype}

\usepackage{inconsolata}

\usepackage{graphicx}

\usepackage{soul}
\newcommand\alley[1]{{\color{magenta} Alley: #1}}

\newcommand\eat[1]{}
\setstcolor{red}

\begin{document}

\date{}

\title{\Large \bf An End-to-End Framework for Functionality-Embedded Provenance Graph Construction and Threat Interpretation}

\author{
{\rm Kushankur Ghosh}\\
University of Alberta\\
kushanku@ualberta.ca
\and
{\rm Mehar Klair}\\
University of Alberta\\
meharsin@ualberta.ca
\and
{\rm Kian Kyars}\\
University of Alberta\\
kkyars@ualberta.ca
\and
{\rm Euijin Choo\thanks{Corresponding author: euijin@ualberta.ca}}\\
University of Alberta\\
euijin@ualberta.ca
\and
{\rm Jörg Sander}\\
University of Alberta\\
jsander@ualberta.ca
} 

\maketitle

\begin{abstract}
\textit{Provenance graphs} model causal system-level interactions from logs, enabling anomaly detectors to learn normal behavior and detect deviations as attacks. However, existing approaches rely on brittle, manually engineered rules to build \textit{provenance graphs}, lack functional context for system entities, and provide limited support for analyst investigation. 
We present \textit{Auto-Prov}, an adaptive, end-to-end framework that leverages large language models (LLMs) to automatically construct \textit{provenance graphs} 
from heterogeneous and evolving logs, 
embed system-level functional attributes into the graph,
enable provenance graph-based anomaly detectors to learn from these enriched graphs, 
and summarize the detected attacks to assist an analyst's investigation.
\textit{Auto-Prov} clusters unseen log types and efficiently extracts provenance edges and 
entity-level information via automatically generated rules.
It further infers system-level functional context 
for both known and previously unseen system entities using a combination of LLM inference and behavior-based estimation.
Attacks detected by provenance-graph-based anomaly detectors trained on \textit{Auto-Prov}'s graphs are then summarized into natural-language text.
We evaluate \textit{Auto-Prov} with four state-of-the-art provenance graph-based detectors across diverse logs. Results show that \textit{Auto-Prov} consistently enhances detection performance, generalizes across heterogeneous log formats, and produces stable, interpretable attack summaries that remain robust under system evolution.
\end{abstract}
\section{Introduction}
\label{sec:intro}
\begin{figure}
\centering
\includegraphics[width=\linewidth,height=10.5cm]{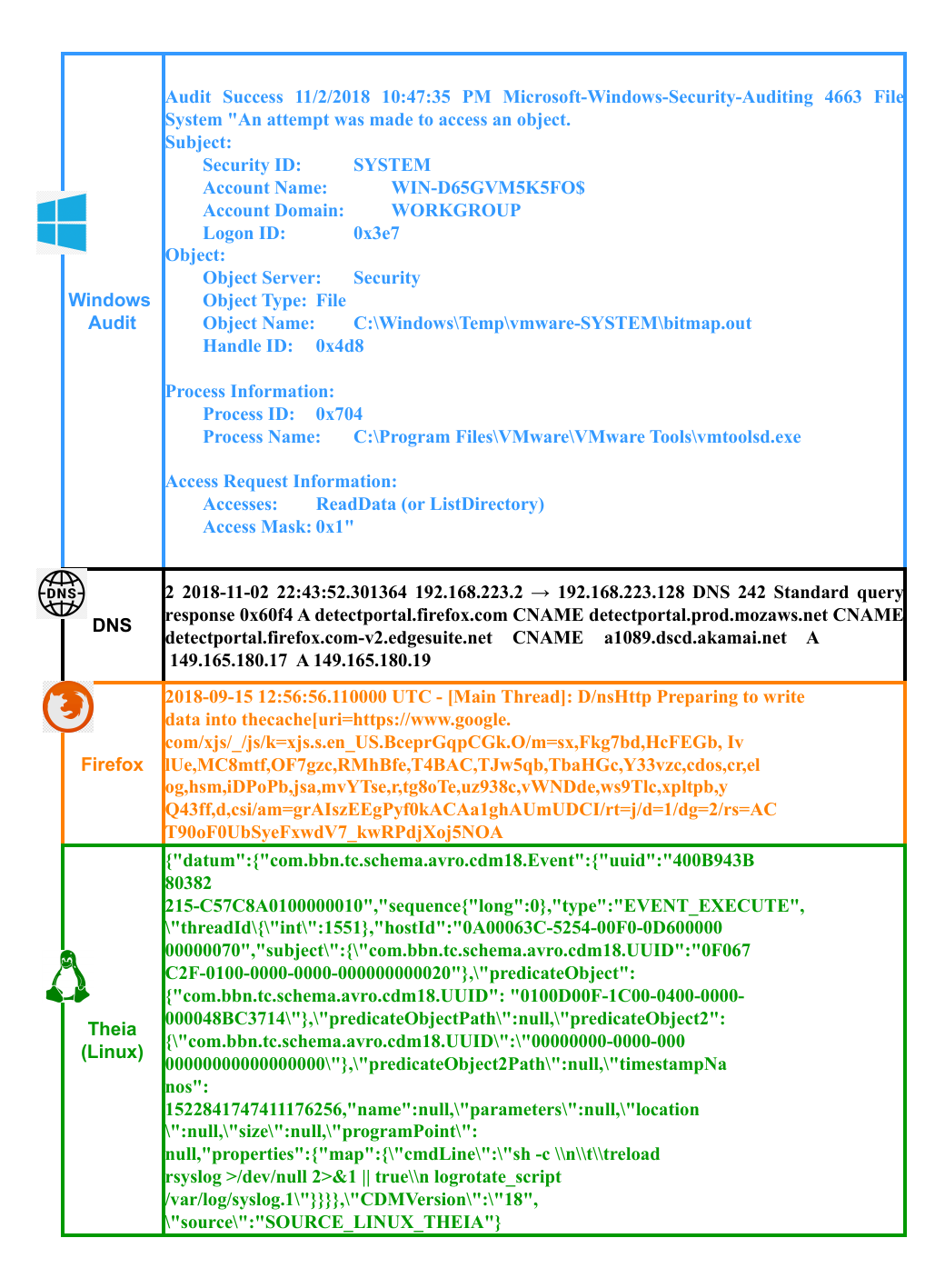}

\caption{Examples of Different Log Entries}
\label{fig:logformats}
\end{figure}

Advanced Persistent Threats (APTs) are
stealthy attacks that evade detection by blending into normal operations and advance slowly through multiple stages \cite{jia2024magic, aly2025ocr}. 
If not detected early, they can establish persistence and escalate privileges, leading to large-scale breaches and data loss.
To spot APT activities, analysts monitor a vast volume of 
logs that capture system-level interactions (e.g., process executions, file accesses, and network connections) to uncover rare or unexpected behaviors.
However, logs are largely unstructured, inconsistent across platforms, and contain low-level details (e.g., \emph{thread ID}), making it difficult to infer meaningful interactions between system entities (e.g., files, processes, and network components) directly from raw logs (see Fig. \ref{fig:logformats}) \cite{du2017deeplog}.

Provenance graphs resolve this challenge by modelling log entries into 
cause-and-effect relationships. In a provenance graph, nodes represent system entities and edges capture causal interactions (e.g., read, write, execute).
By linking independent log entries into a coherent execution flow, provenance graphs enable
anomaly detectors to learn normal behavior built on benign logs and flag deviations as threats \cite{aly2025ocr}.
However, building provenance graphs from raw logs and effectively detecting anomalies 
within them pose challenges.

\noindent \textbf{Challenge 1.} Constructing a provenance graph largely depends on manual effort and does not scale well across heterogeneous and evolving systems.
Existing provenance graph construction approaches rely heavily on manually crafted regular-expression rules to extract entities, their types (e.g., \emph{file, process}), names (e.g., \emph{vmtoolsd.exe}), and interactions from logs \cite{rehman2024flash,cheng2024kairos,jia2024magic}. This process assumes knowledge of all log formats and types. 
Given the volume, heterogeneity, and constant evolution of system logs, crafting and maintaining such rules is labor-intensive. 
As software evolves, even minor changes in log formats can invalidate existing rules, requiring constant manual re-engineering \cite{zhong2024logparser}. 

\noindent \textbf{Challenge 2.} Existing provenance-graph-based anomaly detectors 
struggle to capture the functionality of entities within the graphs due to the limited contextual information available in the logs. 
Prior work typically represents nodes using embeddings of entity names \cite{rehman2024flash, cheng2024kairos} or coarse-grained types (e.g., \textit{process}) \cite{jia2024magic}, often derived from generic embedding models (e.g., \textit{word2vec}) \cite{rehman2024flash}, or through handcrafted behavioral statistics such as action frequencies~\cite{lv2024trec, aly2025ocr}. These representations describe \textit{what} an entity is or \textit{how} it behaves, but not \textit{why} it behaves in that way. Hence, detectors may struggle to learn whether an interaction aligns with entity's functionality, causing benign and malicious behaviors that differ primarily in intent to be modeled similarly and degrading detection performance. For example, a coarse type \emph{process} can encompass functionally distinct subtypes: 
\textit{document editor process} 
calling
a \textit{command-line process} often signals compromise, whereas the same behavior is expected for a \textit{service host process} \cite{wang2020you}.
Similarly, entities with different intents may exhibit comparable behavioral statistics:
malware C2 beaconing and benign system processes such as automated updates may both produce periodic communication patterns \cite{mahboubi2025evolving}.
Also, when using entity names as node features, embedding models trained on general corpora often treat functionally related entities (e.g., \textit{fltmgr.sys} and \textit{storqosflt.sys}) as unrelated, leading to poor detection and limited generalization to previously unseen entities as systems evolve.

\noindent \textbf{Challenge 3.} 
Existing anomaly detectors offer limited support for analyst interpretation. Detected alerts are often presented as large, complex attack graphs that are difficult to interpret and overwhelming in practice \cite{aly2025ocr}. 
\noindent \textbf{Motivation.} 
LLMs have shown promise across a range of tasks involving unstructured data, including constructing cybersecurity knowledge graphs from cyber threat intelligence (CTI) reports \cite{huang2024ctikg,cheng2025ctinexus}, enriching node features to improve graph learning \cite{chen2024exploring}, and generating CTI reports from graphs or logs \cite{aly2025ocr,aldaihan2025clouseau}. 
These studies suggest that LLMs can interpret heterogeneous inputs, infer relationships, extract semantic attributes, and generate higher-level abstractions.
Therefore, LLMs may address \textit{challenges 1–3}; however, directly using them for the provenance graph entails additional challenges.

\textit{First}, 
given the sheer volume of system logs, applying LLMs directly to log streams is computationally impractical.
\textit{Second}, 
constructing graphs from system logs is more challenging than from CTI reports, because logs consist of low-level records that lack the narrative structure, explicit relational cues, paragraphs, sentences, and contextual explanations present in natural-language text, that align closely with the data LLMs are primarily trained on \cite{liu2025datasets}.
\textit{Third},
the use of LLMs to infer functional context for nodes in the provenance graph remains unexplored, where nodes correspond to low-level system entities rather than natural-language descriptions. 
Accurately representing these entities requires inferring their functionality to address \textit{Challenge 2}.
However, many entities in evolving system logs are domain-specific or newly introduced and are often absent from an LLM's pretraining corpus, complicating direct inference.
\textit{Fourth}, 
existing studies that use LLMs to ``summarize'' graphs into reports typically assume that the LLM itself can act as a threat detector, 
relying on embedded security knowledge to identify attack-related artifacts \cite{rehman2024flash, aly2025ocr}.
Such outputs may be driven primarily by the model's prior knowledge of known system files rather than by reasoning over observed system behavior.
In evolving systems containing previously unseen entities, this limits reliability, as LLMs may mischaracterize unfamiliar components without knowledge about their functionalities.


\textbf{Our Approach.} We propose \textit{Auto-Prov}, an adaptive, end-to-end framework 
that automatically models heterogeneous system logs into provenance graphs with functional context for anomaly detection and interpretation.
The goal is to enhance the practicality of provenance-graph-based anomaly detection by making it more resilient to system evolution and bringing it closer to real-world deployment.

\textit{Auto-Prov} 
continuously identifies and clusters distinct log types from log streams, allowing it to accommodate previously unseen log formats (section \ref{subsec:log_clustering}).
On randomly selected samples of each log type, \textit{Auto-Prov} then uses a large LLM to extract the entities (nodes), their associated information (e.g., entity names), and links them by extracting the interaction types between them and timestamps, forming the provenance graph edges (section \ref{subsec:candidate_prov_extractor}).
To scale this process to high-volume log streams, we utilize a smaller open-source LLM to generate rules that replicate the extracted patterns and build the provenance graph (section \ref{subsec:rule_generator}).
\textit{Auto-Prov} then embeds functional context for system entities in the graph. \textit{Auto-Prov} infers the functionality labels (e.g., document editor, browser) by leveraging the open-source LLM for known system entities and a \textit{behavior classifier} for unknown entities(section \ref{subsec:llm_features}). Anomaly detectors trained on the graphs identify malicious nodes and edges, which are linked to form attack graphs.
\textit{Auto-Prov} subsequently summarizes the attack graphs into concise natural-language descriptions and maps to likely APT tactics to assist analysts' investigation (section \ref{subsec:attack_graph_assistant}).\looseness=-1

We evaluate \textit{Auto-Prov} on four state-of-the-art provenance graph-based anomaly detectors \cite{rehman2024flash,jia2024magic,aly2025ocr,cheng2024kairos} having diverse node feature representations. 
Experiments are conducted on heterogeneous logs collected from multiple platforms, which cover six real-world attacks and exhibit substantially different logging formats (Fig. \ref{fig:logformats})~\cite{darpa_tc_e3, alsaheel2021atlas}. We show that \textit{Auto-Prov} (1) successfully identifies different log types; (2) automatically constructs richer provenance graphs from heterogeneous logs; (3) infers functionality labels for known and unknown system entities; (4) consistently improves detection performance of all evaluated detectors; and (5) generates concise natural-language summaries of detected attack graphs that highlight behaviors aligned with APT tactics, and remain consistent with previously unseen entities. Our major contributions are: 
\begin{itemize}
\vspace{-0.5em}
  \setlength{\itemsep}{0pt}   
  \setlength{\parskip}{0pt}   
    \item We propose an adaptive, end-to-end framework, \textit{Auto-Prov}, that automatically constructs provenance graphs from raw logs, infers functional context for system entities, and generates concise attack graph summaries 
    in natural language to support the analyst's investigation.
    \item \textit{Auto-Prov} continuously identifies new log types and leverages LLMs to extract provenance edges and node-level information,
scaling to high-volume log streams. 
    \item We propose a functional-context representation for provenance graph nodes that embeds entities with 
    functionality labels, 
    combining LLM-based inference for known entities with behavior-driven functionality estimation for previously unseen entities. 
    \item We propose an attack summarization approach that translates detected attack graphs into a natural language summary
    and maps observed behaviors to likely APT tactics. 
\end{itemize}

\section{Background and Related Work}
\label{sec:related_work}

\noindent \textbf{Provenance Graph from System Logs.}
System logs record system-level activities by capturing interactions among entities such as processes, files, and network components. 
Individual log entries typically describe isolated events and include entity identifiers, names, coarse entity types, interaction types, and auxiliary metadata such as timestamps \cite{zeng2021watson}.

Provenance graphs model these unstructured, low-level logs as directed graphs, where nodes represent system entities and edges represent causal interactions (e.g., read, write, execute). By linking log entries over time through shared entities, provenance graphs provide a unified execution flow that enables downstream analysis, including anomaly detection \cite{rehman2024flash, jia2024magic, aly2025ocr, cheng2024kairos}. Existing provenance graph-based anomaly detectors construct such graphs using manually crafted, log-specific regular expression rules to extract entities, their associated information, and interactions from raw logs. 
These rules are typically static and must be customized for different platforms, log sources, and datasets, and re-engineered as logging formats evolve. This makes provenance graph construction labor-intensive and difficult to scale in practice.

Prior work on automated log parsing focuses on extracting structured fields or templates from raw logs to facilitate log analysis \cite{du2017deeplog, li2020swisslog,le2023log,ma2024llmparser}. However, these approaches do not address automated provenance graph construction.

\noindent \textbf{Provenance-Graph-based Detection.} Early provenance-graph-based detectors relied on expert-defined attack patterns, making them brittle against evolving attacks \cite{jia2024magic,aly2025ocr,altinisik2023provg, aly2024megr}. Recently, graph neural network-based detectors are used to detect malicious nodes or edges in provenance graphs \cite{rehman2024flash, jia2024magic, aly2025ocr, cheng2024kairos}, but challenges remain. They use static regular expression rules for provenance graph constructions, which become invalid as logs evolve, and use node features 
that fail to capture entity functionality. 
Flash \cite{rehman2024flash} and Kairos \cite{cheng2024kairos} use raw file paths and IP addresses as node features and optimize learning objectives such as predicting coarse entity types \cite{rehman2024flash} or interaction types \cite{cheng2024kairos}.
MAGIC \cite{jia2024magic} use coarse entity and interaction types, and trains to reconstruct masked features and predict edge existence.
OCR-APT \cite{aly2025ocr} use behavioral statistics, such as action frequencies and idle-period features as node features. 
As discussed in section \ref{sec:intro}, these representations makes it challenging for detectors to know whether an observed interaction is expected for a given entity.
Hence, malicious entities resembling the normal interactions of other entities may evade detection. Additionally, most detectors provide limited support for analysts. Alerts are often presented as large, low-level attack graphs that are difficult to interpret.
Recently 
LLMs have been used to generate reports
out of them \cite{aly2025ocr}, but treating LLMs as detectors may be unreliable in evolving systems with newer system entities unknown to the LLM.

\noindent \textbf{LLMs and their Applications.} LLMs have gained attention for strong performance in language understanding and generation tasks, including summarizing and interpreting complex text \cite{fang2024large, ning2024cheatagent, xu2025chatpd}. They exhibit robust contextual understanding and can extract structured information from unstructured texts. 
LLMs typically fall into two categories: proprietary models (e.g., GPT-4o), 
accessed via APIs with restricted interfaces and usage-based costs that can be prohibitive for high-throughput applications \cite{hurst2024gpt, chen2024frugalgpt}; and
\textit{open-source} LLMs (e.g., LLaMA-3), with full parameter access \cite{grattafiori2024llama}. 
While open-source LLMs can be adapted to specific tasks via \textit{fine-tuning}, this often requires substantial labeled data and compute and risks issues such as \textit{catastrophic forgetting}~\cite{li-etal-2024-revisiting}. Alternatively, \textit{in-context learning} allows both model types to adapt at inference time via carefully designed prompts using role specification and \textit{few-shot} \cite{brown2020language} or \textit{zero-shot} \cite{kojima2022large} examples.\looseness=-1


Existing LLM-based work falls into three categories. The first category use LLMs for graph generation. CTINexus \cite{cheng2025ctinexus} extracts entity–relation triplets from CTI reports to build \textit{cybersecurity knowledge graphs}. ProvSyn \cite{huang2025provsyn} and PROVCREATOR \cite{wang2025provcreator} generate synthetic graphs to aid anomaly detection: ProvSyn learns a distribution over existing provenance graphs and synthesizes new ones with topology, labels, and LLM-generated node names, while PROVCREATOR jointly model existing graph structure and node attributes. In contrast, we automatically construct provenance graphs from heterogeneous, evolving system logs in a zero-shot setting.

The second category uses LLMs to enhance node attributes. Prior work generates text-attributed nodes \cite{yu2025leveraging} or pseudo-labels \cite{he2024harnessing, chen2024exploring}, mainly on citation graphs with well-formed text (e.g., titles, abstracts), that align with LLM pretraining data. In contrast, provenance graphs contain low-level system entities (e.g., process) that lack well-formed text, and LLMs may not recognize new or unknown entities appearing as software evolves.

The third category uses LLMs for attack explanation and investigation. Racounteur \cite{deng2024raconteur} explains suspicious shell commands using MITRE ATT\&CK tactics, while Clouseau \cite{aldaihan2025clouseau} and OCR-APT \cite{aly2025ocr} embed LLMs in multi-stage pipelines as detectors of suspicious entities or indicators of compromise (IOCs) to generate reports. 
However, LLM-based detection raises concerns: it is often
unclear whether IOC or suspicious entity detections reflect
reasoning over observed system behavior or biases toward
familiar entity names from the LLM’s pretraining data. LLMs
may mischaracterize unfamiliar entities in evolving systems. 
In contrast, we use LLMs as assistants to summarize detected attack graphs and highlight behaviors consistent with APT tactics to support analysts.



\section{Threat Model}
\label{sec:threat_model}
We adopt a threat model consistent with prior work \cite{rehman2024flash, jia2024magic,cheng2024kairos,jiang2025orthrus}. We assume attackers operate from outside the system and execute stealthy, sophisticated actions that blend into normal system behavior while still leaving observable traces in network- and system-level logs. 
We assume a trusted logging infrastructure. 
 Attackers cannot delete or modify logs once they are generated. Attacks that compromise the logging infrastructure or attempt to poison detectors or LLM components via crafted malicious inputs \cite{chen2025struq} are out of our scope. \looseness-1

\section{Auto-Prov}
\label{sec:autoprov}

\begin{figure*}[t]
\centering
\includegraphics[width=.98\linewidth, trim=0.5cm 0 0 0, clip]{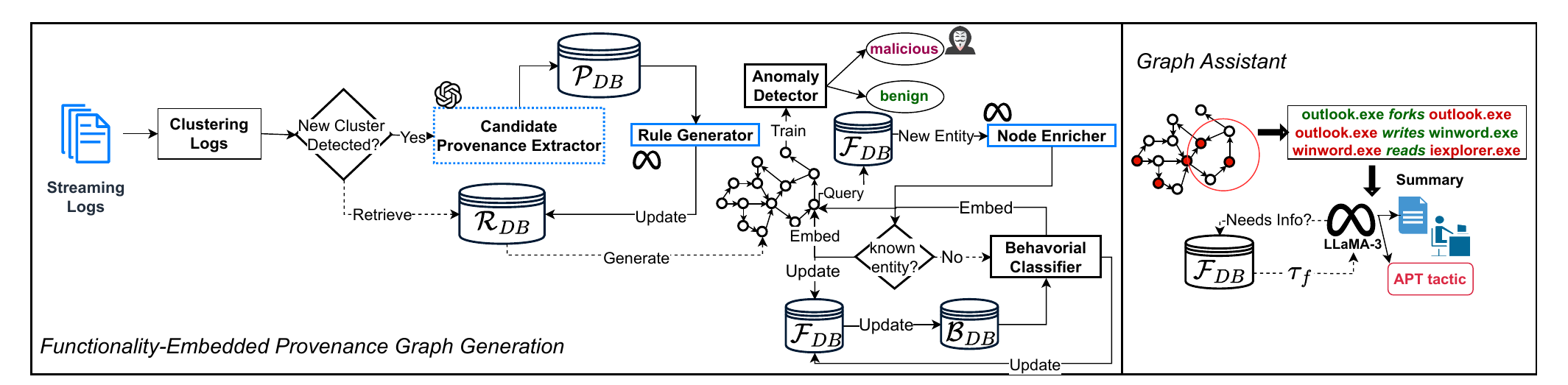}


\caption{Auto-Prov workflow. 
\looseness-1}
\label{fig:auto-prov}
\end{figure*}


\subsection{Overview}
\label{subsec:overview}
{\textit{Auto-Prov} is an end-to-end framework for automatic construction of functionality-embedded provenance graphs to support anomaly detection and attack summarization.
To address three challenges discussed in section \ref{sec:intro}, \textit{Auto-Prov} leverages LLMs in a zero-shot setting. (1) It automatically induces regular expression rules to extract provenance graphs from heterogeneous system logs with evolving structural formats (i.e., the same fields within the log arranged in a consistent pattern) (Challenge 1). 
(2) It enriches graph nodes with inferred functionality attributes, enabling existing provenance graph-based anomaly detectors to better distinguish functionally similar and dissimilar entities (Challenge 2). 
(3) It summarizes detected attack graphs into natural-language descriptions and maps observed events to likely APT tactics 
(Challenge 3). Fig. \ref{fig:auto-prov} depicts \textit{Auto-Prov} consisting of the following components. \looseness=-1


\noindent \textbf{\textcircled{1} Automatic discovery of different log types (section \ref{subsec:log_clustering}).}
To enable scalable provenance graph extraction across heterogeneous and evolving logs, 
\textit{Auto-Prov} exploits the heavy structural redundancy present in system logs. It groups logs sharing a similar structural format into 
the same log type and processes them using a common extraction rule.
By automatically discovering these log types, \textit{Auto-Prov} requires the LLM to process only a small number of representative logs per type to generate extraction rules, which are then applied at scale to the remaining logs.
As log formats are unknown a priori and may evolve over time, \textit{Auto-Prov} discovers log types by clustering continuously arriving logs based on structural similarity and treats each cluster as a distinct log type.
For each discovered type, \textit{Auto-Prov} selects a small set of representative logs, which are aggregated into a \textit{candidate log-set} that collectively covers all observed log formats. \looseness=-1


\noindent \textbf{\textcircled{2} Candidate Provenance Extractor (section \ref{subsec:candidate_prov_extractor}).} 
After identifying representative logs that cover the discovered log formats, \textit{Auto-Prov} requires a set of high-quality provenance extractions to serve as reference points, to generate rules that can be used to replicate those extractions across different log formats at scale. 
These candidate extractions define what information should be extracted from each log type, including interacting entities, interaction type and direction, timestamps, and entity-level information (e.g., names and coarse types). Obtaining such reference provenance information is challenging because raw system logs are dense and unstructured, and inferring which entities interact and how the interactions are directed, therefore, requires substantial reasoning capability. \textit{Auto-Prov} addresses this by applying a large, powerful LLM as the \textit{Candidate Provenance Extractor} (CPE). The CPE operates on the \textit{candidate log-set} and produces candidate provenance graphs, which are used to guide rule generation.

\noindent \textbf{\textcircled{3} Rule Generator (section \ref{subsec:rule_generator})}. 
While the CPE provides high-quality extractions, applying a large LLM to streaming logs is computationally infeasible. 
\textit{Auto-Prov} thus distills the CPE’s extraction behavior into lightweight, reusable rules.
Using each log in the \textit{candidate log-set} and its corresponding CPE output, the \textit{rule generator} induces regular expression rules that reproduce the provenance fields extracted by the CPE (i.e., source and destination entities, interaction types and directions, timestamps, and entity-level information). 
These rules are stored in a \textit{rule database} and are applied to streaming logs to extract the provenance graph
. The \textit{rule database} is incrementally updated whenever a new log type is discovered, allowing \textit{Auto-Prov} to adapt to evolving log formats. \looseness=-1


\noindent \textbf{\textcircled{4} Anomaly Detection with Node Enricher (section \ref{subsec:llm_features}).} To address Challenge 2 (section \ref{sec:intro}), \textit{Auto-Prov} infers \textit{functionality labels} (e.g., document editor, Microsoft driver) to represent nodes using a \textit{node enricher} component. 
When an entity is first encountered, \textit{Auto-Prov} attempts to infer its functionality directly from its name using an LLM. 
To improve efficiency, inferred labels are cached in a \textit{functionality database} and reused for subsequent occurrences of known entities.
When the LLM cannot assign a label\textemdash e.g., for obscure or previously unseen entities\textemdash the \textit{node enricher} falls back to a \textit{behavioral classifier} that estimates labels based on similarity of one-hop system-level interactions to those of known entities.
The inferred labels are embedded as node attributes in the generated provenance graph. \textit{Auto-Prov} then applies existing provenance-graph-based anomaly detectors to the functionality-embedded graph to detect anomalous nodes or edges, which are subsequently linked to form attack graphs representing suspicious activity.


\noindent \textbf{\textcircled{5} Graph Assistant (section \ref{subsec:attack_graph_assistant}).} 
To address the Challenge 3 (section \ref{sec:intro}) and support the human analyst's investigation, \textit{Auto-Prov} converts detected attack graphs into concise, natural-language summaries. The \textit{graph assistant} component of \textit{Auto-Prov} first translates each edge in an attack graph into a natural-language format capturing the source entity, destination entity, and interaction type. These are then provided to an LLM, which generates a summary of the attack graph explaining the sequence of suspicious activities. To further guide analysis, the LLM is instructed to highlight events from the summary and map them to likely APT tactics (e.g., initial compromise), providing reasoning that explains how the characteristics of each event align with the inferred tactic.\looseness=-1

\subsection{Finding Different Log Types}
\label{subsec:log_clustering}

The task of clustering logs to automatically discover different log types presents two key challenges. \textit{First}, system logs are highly redundant. Within short time intervals, millions of entries may share the same format while differing only in variable fields such as timestamps and identifiers. Clustering all such entries is expensive and unnecessary. \textit{Second}, new, previously unseen formats appear over time. Hence, the method needs to continuously assign incoming logs into clusters while simultaneously detecting and incorporating new clusters. \looseness-1

\textit{Auto-Prov} addresses both challenges by (i) sampling logs while preserving format diversity, (ii) clustering sampled logs online to discover newer log types, and (iii) selecting a small set of representative logs per type to characterize each format.

\textbf{Sampling Logs.} \textit{Auto-Prov} samples logs within short time windows (e.g., one hour). 
As \textit{Auto-Prov} samples in successive windows, newly appearing log formats are naturally incorporated. Within each time window $t_F$, every log entry is encoded using RoBERTa \cite{liu2019roberta}, producing an embedding matrix $\mathcal{V}^{t_F}=[v_i]^n_{i=1}$. \textit{Auto-Prov} then selects $k$ maximally dissimilar logs using greedy farthest point (GFP) sampling \cite{li2022adjustable}. GFP iteratively selects the log that maximizes its minimum cosine distance to previously selected samples, yielding a diverse subset of logs $\mathcal{S}^{t_F}$ that spans various formats in that window. The value of $k$ is chosen to be larger than the expected number of log types, ensuring that no distinct format is missed.

\textbf{Clustering Sampled Logs.} Sampling reduces redundancy but does not discover log types. \textit{Auto-Prov} thus clusters the sampled logs $\mathcal{S}^{t_F}$ to group entries with similar formats. To do so,
we use DBStream \cite{hahsler2016clustering}, a density-based clustering algorithm designed for streaming data.
DBStream builds a set of micro-clusters $\mathcal{C}^{t_F}$, each representing a group of log entries whose embeddings lie within a small radius in the embedding space, indicating similar log formats.
Each micro-cluster corresponds to a single log type.
Logs whose embeddings fall outside the radius of existing micro-clusters (i.e., previously unseen formats) are initialized as new clusters. Hence, \textit{Auto-Prov} continuously discovers new log types as formats evolve.\looseness=-1

\textbf{Selecting representative logs per type.} 
As log formats are assumed to be similar within a cluster, we only pass a small number of representative samples to the LLM components to generate the rules. 
For each discovered cluster $c_j \in \mathcal{C}^{t_F}$, \textit{Auto-Prov} randomly selects a 
subset of logs $c'_j \subseteq c_j$ of size $\min(m, |c_j|)$. These logs serve as exemplars of the corresponding format. 
All exemplar logs across clusters are aggregated into a \emph{candidate log-set} $\mathcal{L}^{t_F} = \bigcup_{j=1}^{|\mathcal{C}^{t_F}|}c'_j$, which collectively covers all discovered log formats. This \emph{candidate log-set} is passed to the \textit{candidate provenance extractor} (section \ref{subsec:candidate_prov_extractor}).

\subsection{Candidate Provenance Extractor}
\label{subsec:candidate_prov_extractor}




The \textit{Candidate Provenance Extractor} (CPE) extracts reference provenance graph information directly from raw logs in \textit{candidate log-set} $\mathcal{L}^{t_F}$, including interacting entities, interaction direction and type, timestamps, and available entity-level details. As raw logs are highly unstructured and often omit explicit provenance cues (e.g., source and destination names), this task requires substantial reasoning capability.

A single log entry may contain multiple entity mentions and metadata fields without explicitly indicating which entities participate in the recorded interaction or how they interact. 
Moreover, interaction direction and type are often implicit and need to be inferred. As a result, provenance extraction in a zero-shot setting 
requires resolving several ambiguities.
To do so, the CPE 
leverages a powerful, general-purpose LLM GPT-4o \cite{hurst2024gpt}. To further improve extraction accuracy, CPE decomposes provenance extraction into a sequence of simpler subtasks, following prior work showing that task decomposition improves LLM performance on complex problems \cite{zhou2022least}. Each log entry $l \in \mathcal{L}^{t_F}$ is processed using the following steps. The prompts used for this are in appendix \ref{subsec:cpe_prompts}.

\textbf{Log Summarization.}
\label{subsubsec:LogSummarization}
Logs often contain details irrelevant for provenance graph construction. 
In the first subtask, the LLM generates a natural-language summary describing the system-level interactions captured in $l$.
The model is provided with the raw log entry and minimal contextual information (e.g., operating system).
The summary explicitly captures source and destination entity identifiers (Sid,Did), their names (Sname,Dname), their coarse types (Stype,Dtype) when available, and the interaction type (Itype).
When explicit identifiers are missing, the LLM assigns placeholder identifiers (e.g., entity-1) alongside entity names to enable consistent reference in subsequent subtasks.

\textbf{Entity Types Extraction.}
\label{subsubsec:VanillaEntityTypesExtraction}
Using the generated summary and the log entry $l$, the second subtask extracts the set of distinct entity identifiers appearing in $l$ and assigns each coarse entity type (e.g., process, file) when defined in the log.
The output is a structured mapping $V_l = \{\text{entity-}i:type(\text{entity-}i)\}$.
These coarse entity types are not used as node features in \textit{Auto-Prov}; instead, they are extracted to support learning objectives in certain anomaly detectors (e.g., Flash \cite{rehman2024flash}).

\textbf{Entity Extraction.} 
Not all entities present in a log entry participate in system-level interactions.
For instance, a Windows audit log may include both the executable paths of a process and low-level details, 
only a subset of which correspond to entities participating in the interaction.
This subtask identifies the subset of entities that interact and extracts their names or paths (e.g. \emph{vmtoolsd.exe}) when present.
The LLM first produces a set of interacting entity pairs (entity-$i$, entity-$j$).
For each participating entity, it then extracts the full name or path as it appears in the log, yielding a mapping 
$\mathcal{N}_l = \{\text{entity-}i: name(\text{entity-}i)\}$, which is later used for inferring functional labels (section \ref{subsec:llm_features}).

\textbf{Edge Extraction.}
For each interacting entity pair, the LLM infers the direction of interaction and the interaction type (e.g., read, write).
Given the raw log entry, its summary, and the extracted entity pairs, the model outputs a set of directed edges
$E_l = \{(\text{Sid}, \text{Did}, \text{Itype}, \text{time})\}$, where each edge represents a directed system interaction between two entities.


Processing logs independently may introduce variability, as the LLM can produce multiple plausible outputs for similar entries. To improve consistency, we apply two techniques. 
First, each prompt includes a small, randomly sampled set of prior extractions as in-context examples to encourage consistent conventions across entries \cite{brown2020language}. Second, to reduce random direction inversions during edge extraction, we invoke the LLM multiple times per log entry and apply majority voting over inferred edge directions \cite{wang2022self}. In our experiments, we use seven independent runs and retain the most frequent direction.

After processing each log entry $l\in\mathcal{L}^{t_F}$, the CPE outputs $(E_l, V_l, \mathcal{N}_l)$, 
each capturing a distinct component of the extracted interaction.
These outputs are combined to construct a set of provenance records, 
where each provenance record corresponds to a single directed interaction (i.e., one edge) extracted from $l$.
Formally, each provenance record is defined as 
$p=\{\text{Sid}, \text{Stype}, \text{Sname}, \text{Did}, \text{Dtype}, \text{Dname}, \text{Itype}, \text{time}\}$, 
representing one directed interaction between two entities.
For each $l$, we collect its records into a set $P_l = (p_1, p_2, \ldots, p_k)$. Across the entire \textit{candidate log-set}, this yields the \textit{candidate provenance database}
$\mathcal{P}_{DB} = \{l : P_l\;|\;l \in \mathcal{L}^{t_F}\}$, which serves as the reference input for rule generation (section \ref{subsec:rule_generator}).

\subsection{Rule Generator}
\label{subsec:rule_generator}

For scalability, 
we distill the large LLM's extraction behavior across all the log formats in the \textit{candidate log-set} into a set of reusable regular expression (regex) rules. Given the \textit{candidate provenance database}, $\mathcal{P}_{DB}$, we leverage
a smaller \textit{open-source} LLM (LLaMA-3 \cite{grattafiori2024llama}) 
to induce regex rules that reproduce the field values (e.g., Sid) extracted for each provenance record in $\mathcal{P}_{DB}$. We use the smaller LLM, as rule generation does not require reasoning over unstructured logs. \looseness=-1

The \textit{rule generator} processes each provenance record $p \in P_l$ independently 
and operates in a field-wise manner.
Given a log entry $l$ in the \textit{candidate log-set} and a single field value from $p$ (e.g., $p[$Sid$]$), 
the generator is instructed to induce a regex that extracts the same value from the raw log text.
The per-record output is a rule set 
$r_p$ = $\{ \text{Sid}_r, \text{Stype}_r, \text{Sname}_r, \text{Did}_r, \text{Dtype}_r, \text{Dname}_r, \text{Itype}_r, \text{time}_r\}$.
These rule sets are generated for all provenance records, and unique rules are stored into a \textit{rule database} $\mathcal{R}_{DB}$.


We observe that entity types extracted in section \ref{subsec:candidate_prov_extractor} are often normalized by the LLM (e.g., returning ``file'' when the log contains ``FileObject''). When an extracted type does not appear verbatim in the log, the \textit{rule generator} is instructed to identify the token in the log that best corresponds to the extracted type and to construct a regex that extracts that token. The prompts of this component are in appendix \ref{subsec:rule_generator_prompt}.

Each rule set $r_p \in \mathcal{R}_{DB}$ 
is applied
to raw log entries to extract provenance edges, timestamp, and entity-level information.
Each successful application yields provenance records $p^{(r)}$ of the form $\{$\text{Sid}$,$ \text{Stype}$,$ \text{Sname}$,$ \text{Did}$,$ \text{Dtype}$,$ \text{Dname}$,$ \text{Itype}$,$ \text{time}$\}$, which are linked into a provenance graph.

    

\subsection{Learning with Functional Node Features}
\label{subsec:llm_features}


To address Challenge 2 (section \ref{sec:intro}) and properly capture the system-level functionality of entities,
the \textit{node enricher} component enriches provenance graphs with functional node features that explicitly encode the functionality of each entity.

Each provenance record $p^{(r)}$ from section \ref{subsec:rule_generator} introduces two entities identified by their names Sname and Dname (e.g., file paths, URLs). 
In the constructed provenance graph, each entity name $e$ corresponds to a node $v_e$.
The \textit{node enricher} assigns a functional label $\tau_f(e)$ to each entity
and stores it in a \textit{functionality database} $\mathcal{F}_{DB} = \{e : \tau_f(e)\}$ which ensures consistent functional attribution for the same entity in future logs.\looseness=-1

Functionality label inference begins with name-based reasoning. 
Each $e$ is first normalized to remove incidental variability while preserving semantic structure: redundant path delimiters are removed; numeric substrings (e.g., version numbers) are removed from directory components; and for files, the directory path is sanitized similarly while 
preserving the filename base and extension.
The normalized entity is then passed to an open-source LLM (LLaMA-3 \cite{grattafiori2024llama}), which 
(i) determines whether the input corresponds to a valid system entity (e.g., file, directory, URL),
(ii) if valid, infers its system-level functionality based on naming patterns, path structure, domains, or executable–argument relationships, and
(iii) outputs a functional label $\tau_f(e)$ summarizing the functionality (e.g., document editor, Microsoft driver).
If the LLM cannot infer functionality, 
common for evolving or previously unseen software,
the entity is assigned a special `NO LABEL' tag and deferred to behavior-based inference. The prompts used here are presented in appendix \ref{subsec:node_enricher_prompt}.

For entities with `NO LABEL', \textit{Auto-Prov} 
infers functionality from behavioral signatures observed in the provenance graph.
The underlying intuition is that system-level functionality is reflected in how an entity interacts with other entities: entities with similar functionality tend to participate in similar types of interactions.
For instance, document editors (e.g., MS Word) mostly interact with documents, text files, images, and browsers, whereas system drivers primarily interact with the kernel and hardware interfaces.

For a node $v_e$, we define its \textit{behavioral-profile} $b(v_e)$ as the set of unique one-hop \textit{behavioral signatures} derived from edges adjacent to $v_e$. 
Each signature encodes both the interaction and functional type of the neighboring node, thereby capturing incoming interactions (e.g., read-by web browser) and outgoing interactions (e.g., write-to Microsoft driver).
Let $\mathcal{I}=\{i_1,i_2, \ldots i_d\}$ denote the set of observed \textit{behavioral signatures} across the graph. Each node $v_e$ is represented by a multi-hot vector
$x_{v_e} \in \{0,1\}^d$, where $x_{v_e}[i]=1$ if pattern $s_i \in b(v_e)$, and $0$ otherwise.
\textit{Behavioral profiles} of nodes with known functional labels serve as reference points and are stored in a \textit{behavioral database}
$\mathcal{B}_{DB} =\{ v_e : x_{v_e}\}$. 
For an unlabeled entity $e_{unknown}$, we first construct its multi-hot \textit{behavioral profile} $x_{v_{e_{unknown}}}$ in the same way from its adjacent interactions.
The 
\textit{behavioral classifier} assigns $\tau_f(e_{unknown})$ by identifying the nearest labeled node in this space using cosine similarity and assigning its functional label. 
The inferred label is then added to both the $\mathcal{B}_{DB}$ and $\mathcal{F}_{DB}$, enabling consistent functional attribution for previously unseen entities.


Once the provenance graph is embedded with functional labels, \textit{Auto-Prov} encodes them using MPNet \cite{song2020mpnet}, trains existing provenance graph-based anomaly detectors \cite{rehman2024flash, jia2024magic,aly2025ocr,cheng2024kairos} on benign data to learn normal system behavior. Detected anomalous artifacts (e.g., nodes) are then connected through the provenance graph to construct compact attack graphs, following prior approaches that aggregate anomalies via local graph expansion or community discovery \cite{cheng2024kairos, aly2025ocr}.

\subsection{Attack Graph Assistant}
\label{subsec:attack_graph_assistant}


Addressing Challenge 3 (section \ref{sec:intro}), \textit{Auto-Prov} introduces \textit{graph assistant} based on an open-source LLM (LLaMA-3 \cite{grattafiori2024llama}). Operating strictly as a post-detection layer, it does not perform anomaly detection but converts detected attack graphs into concise, human-readable explanations to aid analyst understanding and guide investigations.



Given a detected attack graph, \textit{Auto-Prov} converts it into a natural-language narrative for LLM-based analysis. Edges are linearized temporally, with nodes resolved to entity names and represented as “source --interaction-type--> target”, where source and target denote entity names. Duplicate consecutive edges are aggregated with a repetition count (e.g., “×10”) to reduce redundancy while preserving behavior intensity.

The graph assistant operates in three stages. 
\textit{First,} it ensures that all entities appearing in the graph are grounded in a functional context. 
For each entity in the attack graph,
the LLM is instructed to assess whether it recognizes the entity’s functionality, producing a binary decision (YES/NO). 
For entities deemed unknown, \textit{Auto-Prov} injects explicit functional context into the prompt.
Specifically, each unknown entity $e_{unknown}$ is paired with its inferred functional label $\tau_f(e_{unknown})$, obtained from the \textit{functionality database} populated by the \textit{behavioral classifier} (section \ref{subsec:llm_features}). These $(e_{unknown}, \tau_f(e_{unknown}))$
tuples are provided as supplementary context in the prompt, while entities recognized by the LLM rely solely on its prior knowledge. This ensures that even previously unseen entity names are anchored to their functionalities, 
enabling coherent reasoning over the attack graph. \looseness=-1


\textit{Second,} the assistant generates a natural-language summary of the attack graph. 
The temporally ordered edge sequence with any injected functional labels for unknown entities is provided to a summarization prompt that instructs the LLM to describe the overall attack behavior captured in the graph. 
The resulting summary highlights the interactions flagged by the anomaly detector and provides context for why they may be suspicious.
An example attack graph and its corresponding summary are shown in appendix \ref{subsec:graph_assistant_outputs}.

\textit{Third,} the assistant further maps the generated attack summary to high-level attacker objectives. 
Given the summary and a set of candidate APT tactics (e.g., Reconnaissance, Privilege Escalation), 
derived from the
MITRE ATT\&CK knowledge base \cite{mitre_attack}, 
the LLM identifies which interactions described in the summary are consistent with each tactic’s defining characteristics and provides a \textit{reasoning} for each mapping. The \textit{reasoning} is a textual explanation of why the observed interactions align with that tactic. The prompts of this component are in appendix \ref{subsec:graph_assistant_prompts}.
This step is not intended to produce authoritative or precise tactic labels. Instead, it offers interpretable hypotheses about the attacker’s possible intent and supporting reasoning to help analysts prioritize alerts and direct their investigation.

\section{Experiments}
\label{sec:evaluation}

We conduct the experiments on a Ubuntu 22.04 server, with AMD EPYC 7713 CPU (64 cores, 2.0 GHz), 503 GB of RAM, and NVIDIA RTX 6000 Ada GPUs with 192 GB memory. We frame our evaluation around the following key questions:\vspace{-0.5em}

\begin{itemize}
\vspace{-0.5em}
  \setlength{\itemsep}{0pt}   
  \setlength{\parskip}{0pt}   
  \item \textbf{RQ1:} Do \textit{Auto-Prov}’s functionality-embedded provenance graphs improve anomaly detection over manually engineered graphs across heterogeneous logs?
  \item \textbf{RQ2:} How do \textit{Auto-Prov}’s individual components contribute to its overall performance?
  \item \textbf{RQ3:} Can \textit{Auto-Prov} summarize attack graphs into accurate and robust natural-language explanations?
  
\end{itemize}

\subsection{Datasets}
\label{subsec:datasets}

We evaluate \textit{Auto-Prov} under realistic conditions characterized by heterogeneous log formats, diverse system platforms, and APT attacks. To ensure coverage along these dimensions, we select datasets that satisfy the following criteria.

\noindent \textbf{(C1) Heterogeneous Log formats.}
\textit{Auto-Prov} is designed to automatically adapt to different log structures without manual engineering. We therefore evaluate it on logs from the widely adopted DARPA-E3 Transparent Computing THEIA dataset \cite{bilot2025sometimes, jiang2025orthrus, darpa_tc_e3} and the ATLAS public datasets (S1–S4) \cite{purseclab_atlas, alsaheel2021atlas, liu2025we,10.5555/3762387.3762567,ding2023airtag,yang2023prographer}. These datasets are collected from various system platforms
, which differ substantially in structure, field semantics, and abstraction level (Fig. \ref{fig:logformats}). THEIA logs are relatively structured and JSON-based, whereas ATLAS logs are less uniform and vary across data sources. Together, they allow us to assess \textit{Auto-Prov}’s ability to operate across heterogeneous log formats.

\noindent \textbf{(C2) Diverse platforms.}
The datasets span multiple platforms and logging infrastructures. THEIA is collected on Ubuntu 12.04 and captures low-level system events such as file accesses and network connections. ATLAS includes Windows 10 audit logs, Firefox logs, and DNS traffic collected. This diversity enables evaluation across fundamentally different system environments.

\noindent \textbf{(C3) Diverse real-world attacks.}
Both datasets contain real-world attacks amid benign background activity (e.g., web browsing, email access). THEIA includes two expert-simulated enterprise attacks: \textit{firefox backdoor w/ drakon in-memory} (FBI) and \textit{browser extension w/ drakon dropper} (BED). ATLAS includes four attacks\textemdash \textit{strategic web compromise} (WEB), \textit{malvertising dominate} (MAD), \textit{spam campaign} (SPAM), and \textit{pony campaign} (PONY)\textemdash covering phishing, malicious code injection, and data exfiltration.


Note that other datasets released under DARPA-E3 or ATLAS do not introduce fundamentally new log formats beyond those considered here. Similarly, log datasets used in prior provenance graph based methods largely follow comparable JSON formats already covered by THEIA~\cite{rehman2024flash, jia2024magic, aly2025ocr, cheng2024kairos}.
Therefore, we believe that the selected datasets provide broad coverage for fairly evaluating \textit{Auto-Prov}’s stability and adaptability across three dimensions we mentioned above. 

\subsection{Setup}
\label{subsec:setup}

\noindent \textbf{Baselines for Provenance Graph Construction.}
\label{subsubsec:baselines}
To enable a fair comparison between manually engineered provenance graphs and those automatically constructed by \textit{Auto-Prov}, we follow standard provenance graph construction practices for each dataset.
For THEIA, we use the provenance graph construction pipelines provided by each considered detector. 
For ATLAS, we use the original provenance graph generator released with the ATLAS dataset \cite{purseclab_atlas}.

\noindent \textbf{Anomaly Detectors.}
To evaluate the effectiveness of \textit{Auto-Prov},
we measure how using functionality-embedded provenance graphs affects the detection performance of state-of-the-art provenance-graph-based anomaly detectors.
We consider four representative systems: Flash \cite{rehman2024flash}, MAGIC \cite{jia2024magic}, OCR-APT \cite{aly2025ocr}, and Kairos \cite{cheng2024kairos}, and use the publicly released source code provided by the respective authors. These are selected to assess whether \textit{Auto-Prov}’s functional labels consistently improve detection across diverse node representations. To the best of our knowledge, we cover all the variants of provenance graph node representation used in the literature.

Note that not all anomaly detectors are applicable to ATLAS when using \textit{Auto-Prov}–generated graphs due to differences in the availability of coarse-grained entity type information. The baseline generator of ATLAS \cite{purseclab_atlas} produces graphs through manual preprocessing steps that explicitly assign coarse entity types (e.g., process, file) to nodes, when such information is absent from the raw logs. However, \textit{Auto-Prov} generates extraction rules from raw logs and extracts entity names, actions, and coarse entity types only when explicitly present in the logs. It does not infer or fabricate missing coarse entity types. As Flash \cite{rehman2024flash} and OCR-APT \cite{aly2025ocr} rely on coarse type information for their training or prediction, we exclude them from our evaluation on ATLAS, while including them in evaluations on THEIA; on the other hand, MAGIC \cite{jia2024magic} and Kairos \cite{cheng2024kairos} do not require entity types for their training or prediction. MAGIC uses coarse entity types only as node features, and Kairos relies on entity names and learns via edge prediction. We thus report results for MAGIC and Kairos on both ATLAS and THEIA.

\noindent \textbf{Ground Truth and Evaluation Measures.}
\label{subsubsec:measures}
For THEIA, we use the ground truth recently released by \cite{bilot2025sometimes, jiang2025orthrus}, 
which provides separate labels for individual attacks, unlike prior work that aggregates all malicious labels \cite{rehman2024flash,jia2024magic,cheng2024kairos,aly2025ocr}. 
This enables attack level analysis.
Also, ground truth annotations in earlier evaluations label many unrelated nodes as malicious, which can distort evaluation \cite{bilot2025sometimes}.
For ATLAS, we obtain the ground truth
from the original data repository~\cite{purseclab_atlas}. 
Following \cite{ding2023airtag}, we map the malicious entities back to the 
corresponding log entries, labeled as malicious. 


Prior work primarily reported threshold-based metrics such as precision, recall, F1, or accuracy, which are evaluated at a single decision threshold and can be sensitive to threshold selection, potentially biasing results~\cite{bilot2025sometimes}. In contrast, AUC-ROC (Area Under ROC curve) and AUC-PR (Area Under Precision-Recall Curve) consider the full distribution of anomaly scores. AUC-ROC measures the probability that an attack instance is assigned a higher anomaly score than a benign instance, while AUC-PR emphasizes ranking quality when attack instances are rare by focusing on precision among the highest-scored nodes. Therefore, we use them in our investigation.

Recent research pointed out a challenge in evaluating provenance graph-based anomaly detectors using threshhold-based or ranking-based metrics, and proposed a new metric, called Attack Detection Precision (ADP)~\cite{bilot2025sometimes}. Concretely, some attacks produce large anomalous subgraphs that are relatively easy to detect, while others are stealthier and manifest as only a few anomalous nodes. Threshold-based or ranking-based metrics can therefore be dominated by attacks with many anomalous nodes, masking failures to detect smaller or stealthier attacks. To better assess whether detectors identify all attacks, we thus report ADP~\cite{bilot2025sometimes}. ADP evaluates detection performance at the attack level by quantifying, across precision budgets, the fraction of attacks for which at least one attack node is correctly ranked.

\subsection{Detection Performance}
\label{subsec:detection_acc}

Tables \ref{tab:detection_theia} and \ref{tab:detection_atlas} report detection performance on the THEIA and ATLAS across evaluated anomaly detectors.  
For each detector, ($\beta$) 
denotes the results obtained when training the anomaly detectors using the baseline provenance graphs, 
while Auto-Prov denotes the results when training on our functionality-embedded provenance graphs.

\begin{table*}
\centering
\resizebox{18cm}{!}{
\begin{tabular}{c|cccc|cccc|cccc|cccc} 
\toprule
\multirow{2}{*}{\begin{tabular}[c]{@{}c@{}}\textbf{Provenance}\\\textbf{Framework}\end{tabular}} & \multicolumn{4}{c|}{\textbf{Flash \cite{rehman2024flash}}}                                                                           & \multicolumn{4}{c|}{\textbf{MAGIC \cite{jia2024magic}}}                                                                                     & \multicolumn{4}{c|}{\textbf{OCR-APT \cite{aly2025ocr}}}                                                                                 & \multicolumn{4}{c}{\textbf{Kairos \cite{cheng2024kairos}}}                                                                                       \\
                                                                                                 & \textbf{\textbf{Attack}} & \textbf{AUC-ROC}        & \textbf{AUC-PR}          & \textbf{ADP}                  & \textbf{\textbf{\textbf{\textbf{Attack}}}} & \textbf{AUC-ROC}        & \textbf{AUC-PR}         & \textbf{ADP}           & \textbf{\textbf{\textbf{\textbf{Attack}}}} & \textbf{AUC-ROC}        & \textbf{AUC-PR}         & \textbf{ADP}         & \textbf{\textbf{\textbf{\textbf{Attack}}}} & \textbf{AUC-ROC}        & \textbf{AUC-PR}          & \textbf{ADP}            \\ 
\hline
\multirow{3}{*}{($\beta$)}                                                                             & FBI                      & 0.666                   & 0.0000                   & \multirow{3}{*}{0.5}          & FBI                                        & 0.593                   & 0.027                   & \multirow{3}{*}{0.507} & FBI                                        & 0.47                    & 0.0000                  & \multirow{3}{*}{1.0} & FBI                                        & 0.803                   & 0.0027                    & \multirow{3}{*}{0.666}  \\
                                                                                                 & BED                      & 0.671                   & 0.0001                   &                               & BED                                        & 0.79                    & 0.0013                  &                        & BED                                        & 0.65                    & 0.0001                  &                      & BED                                        & 0.806                   & 0.0008                   &                         \\
                                                                                                 & \textit{\textbf{Avg.}}   & \textit{0.668}          & \textit{0.0000}          &                               & \textit{\textbf{\textbf{Avg.}}}            & 0.\textit{691}          & \textit{0.014}          &                        & \textit{\textbf{\textbf{Avg.}}}            & \textit{0.56}           & 0.0000                  &                      & \textit{\textbf{\textbf{Avg.}}}            & \textit{0.804}          & \textit{\textbf{0.0014}} &                         \\ 
\hline
\multirow{3}{*}{Auto-Prov}                                                             & FBI                      & 0.901                   & 0.001                   & \multirow{3}{*}{\textbf{1.0}} & FBI                                        & 0.81                    & 0.015                   & \multirow{3}{*}{\textbf{1.0}}   & FBI                                        & 0.962                   & 0.0036                  & \multirow{3}{*}{1.0} & FBI                                        & 0.858                   & 0.0053                    & \multirow{3}{*}{0.666}  \\
                                                                                                 & BED                      & 0.853                   & 0.001                    &                               & BED                                        & 0.869                   & 0.017                   &                        & BED                                        & 0.826                   & 0.006                   &                      & BED                                        & 0.813                   & 0.0001                   &                         \\
                                                                                                 & \textbf{\textit{Avg.}}   & \textbf{\textit{0.877}} & \textbf{\textit{0.001}} &                               & \textit{\textbf{\textbf{Avg.}}}            & \textit{\textbf{0.839}} & \textit{\textbf{0.016}} &                        & \textit{\textbf{\textbf{Avg.}}}            & \textit{\textbf{0.894}} & \textit{\textbf{0.005}} &                      & \textit{\textbf{\textbf{Avg.}}}            & \textit{\textbf{0.835}} & \textit{\textbf{0.003}}           &                         \\
\bottomrule
\end{tabular}}
\caption{Detection results on the THEIA dataset \cite{darpa_tc_e3}.}
\label{tab:detection_theia}
\end{table*}

\noindent \textit{\textbf{Finding 1.} Auto-Prov generalizes across diverse log formats.} 



\noindent Tables \ref{tab:detection_theia} and \ref{tab:detection_atlas} show that \textit{Auto-Prov} consistently improves detection performance across datasets with fundamentally different log structures. Notably, \textit{Auto-Prov} improves average AUC-ROC by 0.23 on THEIA and 0.15 on ATLAS across all applicable detectors. We significantly improve ADP for Flash on THEIA (0.5) and MAGIC on THEIA (0.493) and ATLAS (0.61). Importantly, these improvements are observed despite ATLAS logs being substantially less structured.


Note that baseline AUC-PR values are often near zero, as reported in recent work~\cite{bilot2025sometimes}, reflecting class imbalance and stealthy attacks with sparse manifestations. 
In this setting, even modest absolute gains (e.g., 0.002) in AUC-PR are meaningful. \textit{Auto-Prov}'s consistent improvements indicate better prioritization of malicious entities among top-ranked alerts, while larger AUC-ROC gains indicate improved global separation between benign and malicious behavior.




    
\noindent \textit{\textbf{Finding 2.} Functionality features and provenance graphs generated by \textit{Auto-Prov} outperform existing feature representations and manually-constructed graphs.}

\noindent \textit{Auto-Prov}’s functionality-embedded provenance graphs consistently improve AUC-ROC over baselines, indicating better separation between benign and malicious behavior. The largest improvements occur for OCR-APT across diverse attacks, suggesting that functionality features capture richer context than behavioral statistics. 
The overall AUC-PR gains are smaller but consistent, reflecting reduced false positives while remaining challenged by attacks that mimic benign activity. Importantly, \textit{Auto-Prov}’s goal is not to maximize node-level precision, but to provide anomaly detectors with richer provenance graphs that reliably surface attack instances for analyst investigation. This is reflected in the consistently high ADP.\looseness-1

\noindent \textit{\textbf{Finding 3.} Auto-Prov enables reliable attack-level detection across diverse attacks.}



\noindent Across all datasets and detectors, detectors trained on \textit{Auto-Prov}-generated graphs 
often reaches $1.0$ ADP and never degrades relative to baseline graphs. 
This is because \textit{Auto-Prov} increases the likelihood that at least one attack-related node is prioritized among the top alerts,
including stealthy attacks generating only a few anomalous events. 
In practice, such results are often more valuable than improvements in aggregated ranking metrics, 
which ensures that attacks are rarely missed.\looseness-1

\begin{table}
\centering
\resizebox{8cm}{!}{
\begin{tabular}{c|cccc|cccc} 
\toprule
\multirow{2}{*}{\begin{tabular}[c]{@{}c@{}}\textbf{Provenance}\\\textbf{Framework}\end{tabular}} & \multicolumn{4}{c|}{\textbf{Kairos \cite{cheng2024kairos}}}                                                              & \multicolumn{4}{c}{\textbf{MAGIC \cite{jia2024magic}}}                                                                                \\
                                                                                                 & \textbf{Attack}        & \textbf{AUC-ROC}        & \textbf{AUC-PR}         & \textbf{ADP}         & \textbf{\textbf{Attack}} & \textbf{\textbf{AUC-ROC}} & \textbf{AUC-PR}         & \textbf{ADP}                     \\ 
\hline
\multirow{5}{*}{($\beta$)}                                                                             & MAD                    & 0.498                   & 0.024                   & \multirow{5}{*}{0.997} & MAD                      & 0.5                       & 0.046                   & \multirow{5}{*}{0.385}           \\
                                                                                                 & PONY                   & 0.498                   & 0.024                   &                      & PONY                     & 0.498                     & 0.138                   &                                  \\
                                                                                                 & SPAM                   & 0.504                   & 0.008                   &                      & SPAM                     & 0.503                     & 0.044                   &                                  \\
                                                                                                 & WEB                    & 0.498                   & 0.007                   &                      & WEB                      & 0.499                     & 0.052                   &                                  \\ 
\cline{2-4}\cline{6-8}
                                                                                                 & \textit{\textbf{Avg.}} & \textit{0.499}          & \textit{0.015}          &                      & \textbf{\textit{Avg.}}   & \textit{0.5}              & \textit{0.07}           &                                  \\ 
\hline
\multirow{5}{*}{Auto-Prov}                                                             & MAD                    & 0.683                   & 0.037                   & \multirow{5}{*}{0.997} & MAD                      & 0.815                     & 0.132                   & \multirow{5}{*}{\textbf{0.995}}  \\
                                                                                                 & PONY                   & 0.622                   & 0.031                    &                      & PONY                     & 0.64                      & 0.202                   &                                  \\
                                                                                                 & SPAM                   & 0.625                   & 0.01                    &                      & SPAM                     & 0.692                     & 0.08                    &                                  \\
                                                                                                 & WEB                    & 0.606                   & 0.009                   &                      & WEB                      & 0.526                     & 0.067                   &                                  \\ 
\cline{2-4}\cline{6-8}
                                                                                                 & \textbf{\textit{Avg.}} & \textbf{\textit{0.634}} & \textbf{\textit{0.021}} &                      & \textbf{\textit{Avg.}}   & \textit{\textbf{0.668}}   & \textit{\textbf{0.12}} &                                  \\
\bottomrule
\end{tabular}}

\caption{Detection results on ATLAS datasets \cite{alsaheel2021atlas}.}
\label{tab:detection_atlas}
\end{table}

\begin{figure*}[t]
\centering
\begin{subfigure}{3.2cm}
\centering
\includegraphics[width=3.2cm]{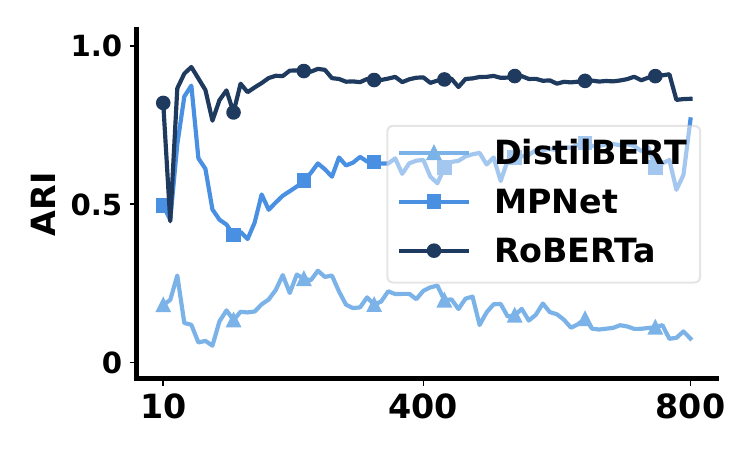}
\caption{}
\label{subfig:clustering_embed_perf}
\end{subfigure}
\begin{subfigure}{3.2cm}
\centering
\includegraphics[width=3.2cm]{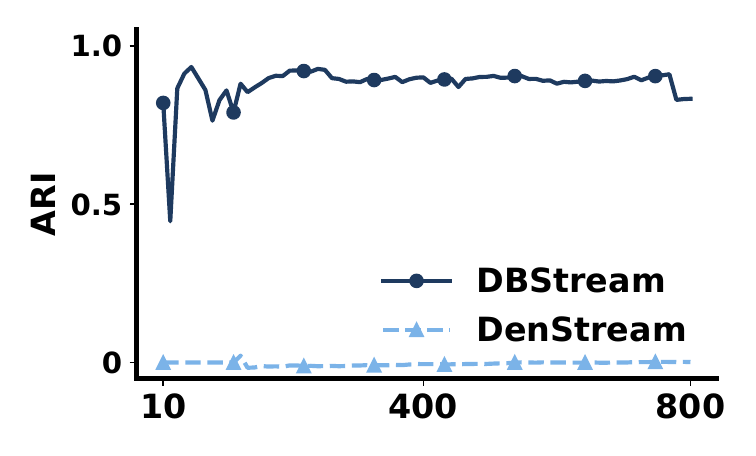}
\caption{}
\label{subfig:clustering_methods_perf}
\end{subfigure}
\begin{subfigure}{3.2cm}
\centering
\includegraphics[width=3.2cm]{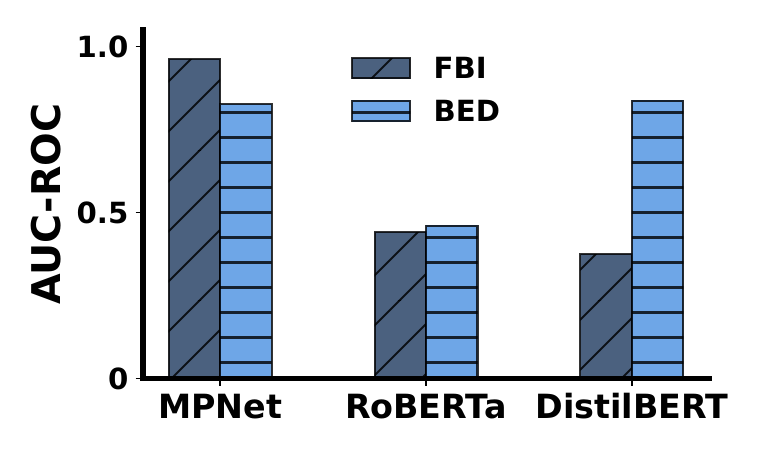}
\caption{}
\label{subfig:detection_embed_perf}
\end{subfigure}
\begin{subfigure}{3.2cm}
\centering
\includegraphics[width=3.2cm]{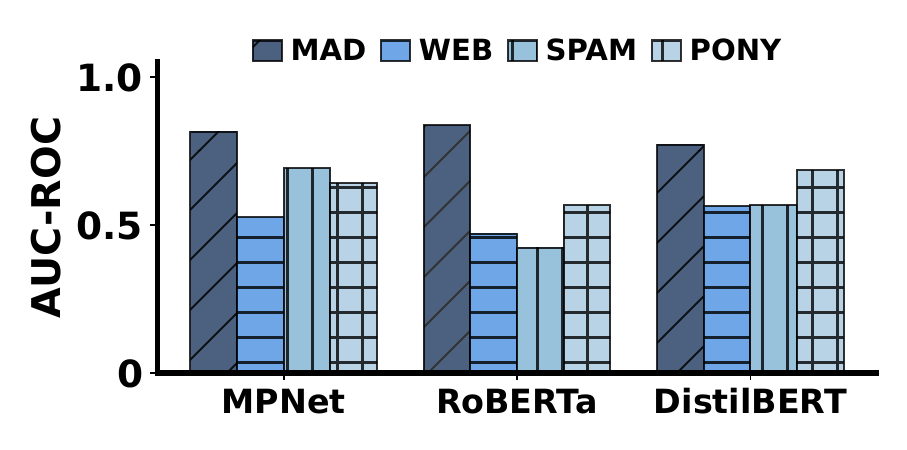}
\caption{}
\label{subfig:detection_embed_perf_atlas}
\end{subfigure}
\begin{subfigure}{4.65cm}
\centering
\includegraphics[width=4.5cm]{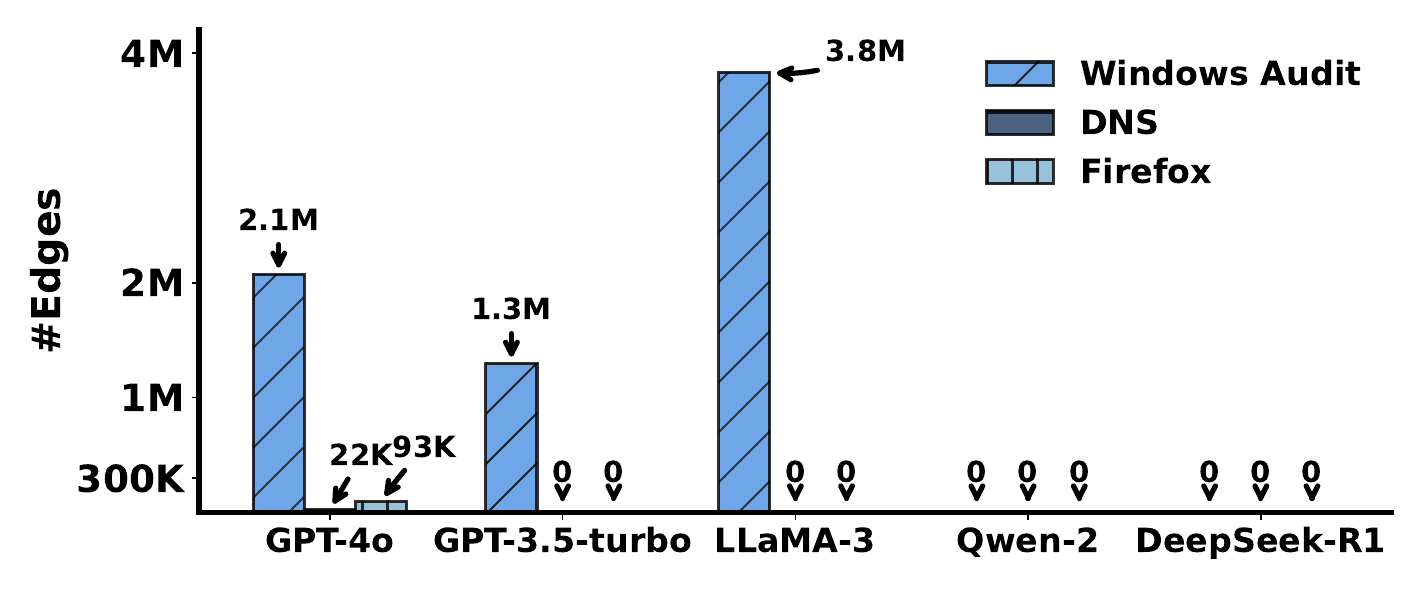}
\caption{}
\label{subfig:edge_number_CPE}
\end{subfigure}

\caption{\textbf{(a):} Clustering performance of DBStream \cite{hahsler2016clustering} across different log embedding models. \textbf{(b):} Clustering performance of DBStream \cite{hahsler2016clustering} and DenStream \cite{cao2006density}. \textbf{(c)}–\textbf{(d):} Detection performance on THEIA \cite{darpa_tc_e3} and ATLAS \cite{alsaheel2021atlas} with different embedding models for node features, respectively. \textbf{(e):} Number of edges generated by different LLMs as the \textit{candidate provenance extractor}.}
\label{fig:clustering_embedding_detection}
\end{figure*}

\subsection{Framework Analysis}
\label{subsec:framework_analysis}

\subsubsection{Log Clustering}
\label{subsubsec:log_clustering}


\textit{Auto-Prov} clusters streaming logs online 
to discover heterogeneous log structures as new log types emerge, enabling downstream provenance extraction without prior knowledge of log formats. 
We evaluate \textit{Auto-Prov}’s design choices by comparing alternative streaming clustering algorithms and log embedding models.


\noindent \textbf{Streaming logs simulation.} To approximate a realistic streaming setting, we construct an incremental log stream from the THEIA.
THEIA logs provide explicit indicators of log type (e.g., Event, FileObject), which we use solely to establish ground truth for evaluation. In practice, we may not have such indicators, and the clustering algorithm is expected to discover them automatically. THEIA logs are assigned to their corresponding log types based on the indicators. Each log type differs in structure and semantics.

Logs are grouped by type 
and randomly sampled to form balanced ground-truth clusters. The grouped logs are shuffled and randomly given incrementally to the clustering. Then, 
a log is drawn from a randomly selected group and processed online. This process continues until all logs are exhausted, yielding a controlled yet realistic simulation of streaming logs. \looseness=-1


\begin{table*}
\centering
\resizebox{18cm}{!}{
\begin{tabular}{c|cc|cc|cc|cc|cc|cc} 
\toprule
\multirow{3}{*}{\textbf{Configuration}} & \multicolumn{4}{c|}{\textbf{THEIA \cite{darpa_tc_e3} + OCR-APT \cite{aly2025ocr}}}                         & \multicolumn{8}{c}{\textbf{ATLAS \cite{alsaheel2021atlas} + Kairos \cite{cheng2024kairos}}}                                                                                                     \\
                                        & \multicolumn{2}{c|}{\textbf{FBI}} & \multicolumn{2}{c|}{\textbf{BED}} & \multicolumn{2}{c|}{\textbf{MAD}} & \multicolumn{2}{c|}{\textbf{PONY}} & \multicolumn{2}{c|}{\textbf{SPAM}} & \multicolumn{2}{c}{\textbf{WEB}}  \\ 
\cline{2-13}
                                        & AUC-ROC & AUC-PR                  & AUC-ROC & AUC-PR                  & AUC-ROC & AUC-PR                  & AUC-ROC & AUC-PR                   & AUC-ROC & AUC-PR                   & AUC-ROC & AUC-PR                  \\ 
\hline
($\beta$) + $\tau_f$                     & 0.48    & 0.0000                  & 0.344   & 0.0001                  & 0.5     & 0.024                   & 0.5     & 0.024                    & 0.503   & 0.08                     & 0.5     & 0.007                   \\
Auto-Prov - $\tau_f$               & 0.532   & 0.0007                  & 0.7     & 0.0001                  & 0.529   & 0.022                   & 0.544   & 0.023                    & 0.52    & 0.007                    & 0.514   & 0.006                   \\
Auto-Prov                               & 0.962   & 0.0036                  & 0.826   & 0.006                   & 0.683   & 0.037                   & 0.622   & 0.031                     & 0.625   & 0.01                     & 0.606   & 0.009                   \\
\bottomrule
\end{tabular}}
\caption{Detection performance comparison between Auto-Prov and its variants: (1) ``($\beta$) + $\tau_f$'': baseline provenance graphs with Auto-Prov’s functionality features; (2) ``Auto-Prov - $\tau_f$'': Auto-Prov-generated provenance graphs with the baseline node features.\looseness-1}
\label{tab:ablation}
\end{table*}

\noindent \textbf{Clustering algorithms and evaluation measures.} 
We evaluate clustering quality using the Adjusted Rand Index (ARI), which measures pairwise agreement between predicted clusters and ground truth~\cite{hubert1985comparing, teitz2023potential, mani2025securing}. 
We compare two streaming clustering methods: DBStream \cite{hahsler2016clustering}, used in \textit{Auto-Prov}, and DenStream \cite{cao2006density}. 
Both operate online and require no predefined number of clusters.
We use the implementations in \cite{montiel2021river} with default hyperparameters and report ARI at each iteration.

\noindent \textit{\textbf{Finding 4.} DBStream more reliably distinguishes logs than DenStream.}


\noindent Fig. \ref{subfig:clustering_methods_perf} shows that DBStream consistently achieves high clustering quality throughout the stream, maintaining an ARI above $0.8$ across iterations. In contrast, DenStream yields ARI values close to zero.
These results justify our choice of DBStream for online log clustering in \textit{Auto-Prov}. 

\noindent \textbf{Log embeddings.} 
To study the impact of log embeddings on clustering performance, we compare RoBERTa \cite{liu2019roberta} used in \textit{Auto-Prov} with two widely used embedding models, MPNet \cite{song2020mpnet} and DistilBERT \cite{sanh2019distilbert}, for encoding raw streaming logs. 


\noindent \textit{\textbf{Finding 5.} RoBERTa-based embeddings provide the most stable representation for streaming log clustering.}



\noindent Fig. \ref{subfig:clustering_embed_perf} shows that RoBERTa consistently outperforms MPNet and DistilBERT in ARI, indicating better capture of structural and semantic log distinctions and more reliable online clustering. Since accurate log-type discovery impacts downstream provenance extraction and rule generation, these results motivate our use of DBStream with RoBERTa in \textit{Auto-Prov}.


\subsubsection{Ablation Studies}
\label{subsubsec:ablation}

\noindent \textbf{Auto-Prov Components.} We evaluate whether \textit{Auto-Prov}’s gains stem from richer graphs generated by \textit{Auto-Prov}, functionality features, or their combination.
`($\beta$) + $\tau_f$' denotes the baseline graph with our functionality features, and `Auto-Prov - $\tau_f$' denote \textit{Auto-Prov} without the functionality features. We evaluate configurations only when applicable.
For THEIA, we use OCR-APT, which shows the largest performance gains in section \ref{subsec:detection_acc}. 
For ATLAS, we use Kairos, as it operates directly on entity names and edges. 
We exclude MAGIC from this analysis because it relies on coarse entity types, 
and hence, MAGIC cannot be applied to the `Auto-Prov - $\tau_f$'.

\noindent \textit{\textbf{Finding 6.} Auto-Prov’s graphs capture more meaningful system relationships than manually engineered rules.}


\noindent Comparing Auto-Prov - $\tau_f$ with the baseline results in Tables  \ref{tab:ablation} and  \ref{tab:detection_theia} shows 
consistent performance improvements even without functional features.
This indicates that \textit{Auto-Prov}-generated graphs model higher-fidelity causal relationships than rule-based pipelines.
Manually engineered regex rules rely on predefined assumptions and often miss valid dependencies, whereas \textit{Auto-Prov} leverages LLM-based reasoning to model plausible interactions directly from logs.

\noindent \textit{\textbf{Finding 7.} Functional features amplify Auto-Prov’s graphs but cannot compensate for deficiencies in baseline graphs.}



\noindent \textit{Auto-Prov}-generated graphs with functionality features yield substantial detection gains, demonstrating that functional context complements the graph. In contrast, embedding functionality features into baseline graphs provides limited or even negative effects. For example, for the BED attack, adding functionality to the baseline graph reduces AUC-ROC from 0.65 to 0.34 (Table \ref{tab:ablation}). This suggests that when the underlying provenance graph is incomplete, functional context alone cannot correct structural errors and may instead amplify noise.

\begin{figure*}[t]
\centering

\begin{subfigure}{3.2cm}
\centering
\includegraphics[width=3.2cm]{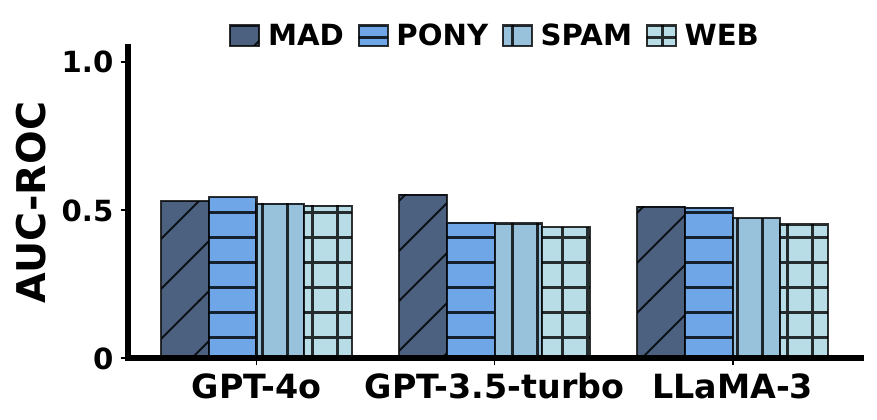}
\caption{}
\label{subfig:CPE_comparison}
\end{subfigure}
\begin{subfigure}{3.2cm}
\centering
\includegraphics[width=3.2cm]{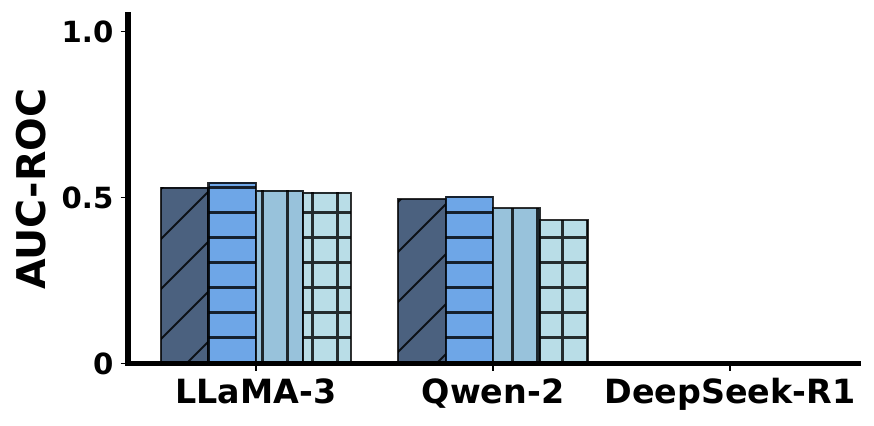}
\caption{}
\label{subfig:rule_generator_comparison}
\end{subfigure}
\centering
\begin{subfigure}{3.2cm}
\centering
\includegraphics[width=3.2cm]{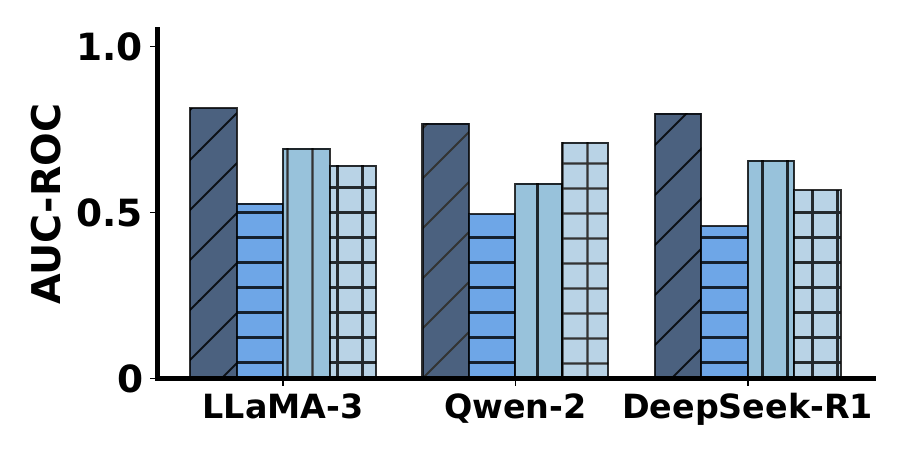}
\caption{}
\label{subfig:llm_nodeenricher_comparison}
\end{subfigure}
\begin{subfigure}{3.2cm}
\centering
\includegraphics[width=3.2cm]{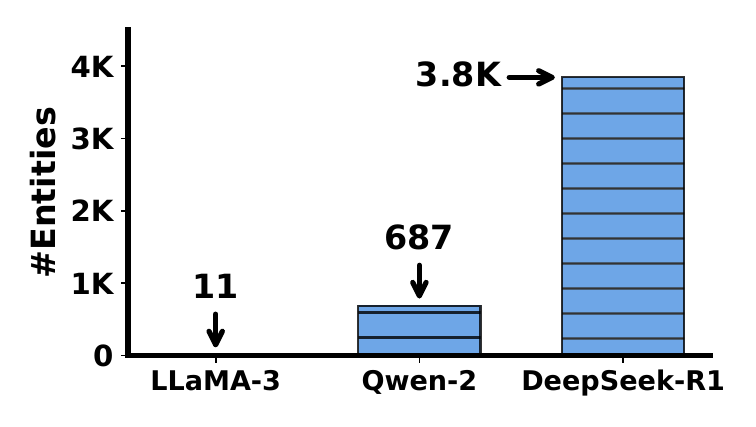}
\caption{}
\label{subfig:unclassifiedEntity_comparison}
\end{subfigure}
\begin{subfigure}{3.2cm}
\centering
\includegraphics[width=3.2cm]{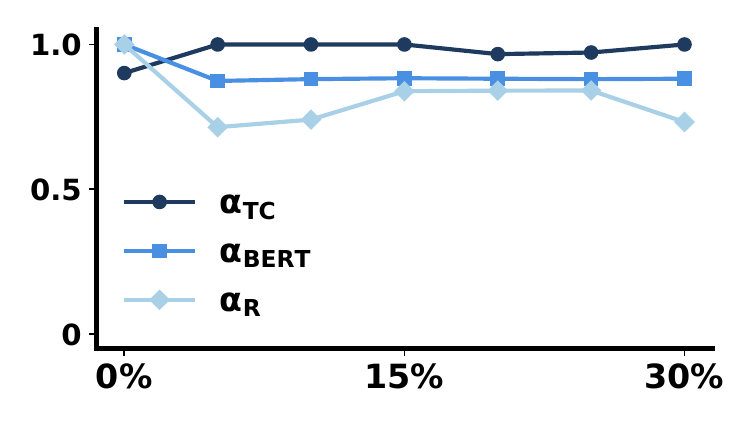}
\caption{}
\label{subfig:poison}
\end{subfigure}


\caption{\textbf{(a)}–\textbf{(c):} Detection performance with different LLMs as \emph{candidate provenance extractor}, \emph{rule generator}, and \textit{node enricher}, respectively. \textbf{(d):} Number of entities for which functionality features cannot be inferred due to insufficient LLM knowledge. \textbf{(e):} Assistant robustness under entity-name poisoning, measured by \textit{tactic correctness} $\alpha_{TC}$, summary similarity $\alpha_{BERT}$, and \textit{tactic consistency} $\alpha_{R}$, averaged over attacks in THEIA \cite{darpa_tc_e3} and ATLAS \cite{alsaheel2021atlas} at varying poisoning rates.}
\label{fig:llms_CPE_rulegenerator}
\end{figure*}


\noindent \textbf{LLMs for Candidate Provenance Extractor.} 
We evaluate multiple LLMs\textemdash GPT-4o \cite{hurst2024gpt}, GPT-3.5-turbo \cite{openai2022chatgpt}, LLaMA-3-70B \cite{grattafiori2024llama}, Qwen-2-72B \cite{team2024qwen2}, and DeepSeek-R1-32B \cite{guo2025deepseek}\textemdash as \textit{candidate provenance extractors} using ATLAS logs. The prompt is held constant across models. 
We use ATLAS for this study, as its logs are highly heterogeneous and substantially less structured than those in THEIA. 
To isolate provenance graph construction quality, functionality features are disabled, and we measure AUC-ROC using Kairos, which operates directly on entity names and graph structure.

\noindent \textit{\textbf{Finding 8.} GPT-4o recovers provenance structure more consistently across heterogeneous logs.}


\noindent 
Fig. \ref{subfig:CPE_comparison} shows that Kairos trained on GPT-4o-generated graphs achieves higher AUC-ROC across attacks compared to graphs produced by other LLMs. While absolute AUC-ROC values remain moderate, the relative differences reflect variations in structural completeness and cross-log connectivity, rather than downstream detection capacity. GPT-4o preserves a broader set of inter-entity relations that enable Kairos to surface attack-relevant deviations that are otherwise weakly observable.

\noindent \textit{\textbf{Finding 9.} Smaller LLMs struggle to recover cross-log provenance interactions.}

\noindent Fig. \ref{subfig:edge_number_CPE} shows that GPT-4o extracts candidate graphs spanning Windows audit, DNS, and Firefox logs. In contrast, GPT-3.5-turbo and LLaMA-3 primarily recover from Windows audit level, 
while Qwen-2 and DeepSeek-R1 fail to generate candidates.
AUC-ROC alone may miss these differences, as many attacks manifest only partially at the audit level.
Even with richer cross-log provenance, overall detection remains bounded  by attacks with sparse observable behavior.
Detectors trained on 
GPT-3.5-turbo and LLaMA-3 graphs
are 
limited to audit activity, with little to no visibility into network or browser behavior.
However, GPT-4o’s ability to retain cross-log interactions becomes increasingly important for attacks dominated by network or browser activity \cite{gast2024snailload}.

\noindent \textit{\textbf{Finding 10.} Smaller LLMs require explicit anchors to extract provenance interactions.}



\noindent GPT-3.5-turbo and LLaMA-3 succeed primarily when logs contain explicit anchors (e.g., process name). They fail on highly unstructured logs (e.g., DNS), where relationships must be inferred, not directly stated. This highlights the need for large, general-purpose LLMs with strong reasoning capabilities for provenance extraction across heterogeneous logs.

\noindent \textbf{LLMs for Rule Generator.} 
Building on the previous investigation, we evaluate open-source LLMs as \textit{rule generators} (Fig. \ref{subfig:rule_generator_comparison}). We compare LLaMA-3-70B \cite{grattafiori2024llama}, Qwen-2-72B \cite{team2024qwen2}, and DeepSeek-R1-32B \cite{guo2025deepseek}, using Kairos on the ATLAS. As before, the prompt is held constant across all models to isolate the effect of rule generation quality.

\noindent \textit{\textbf{Finding 11.} LLaMA-3 generates the most complete rule sets across heterogeneous logs.}


\noindent Regular expression rules generated by LLaMA-3 yield the highest AUC-ROC across all evaluated attacks. Qwen-2 achieves comparable performance on several attacks but occasionally fails to generate rules for specific HTTP interactions in Firefox logs, leading to modest performance degradation. 
In contrast, DeepSeek-R1 fails to generate usable rules, consistent with its inability to act as an effective candidate provenance extractor in the prior evaluation. 

    

\noindent \textbf{Node Embedding Models.}  
Fig. \ref{subfig:detection_embed_perf} and Fig. \ref{subfig:detection_embed_perf_atlas} evaluate the impact of different node embedding models on detection performance when trained on \textit{Auto-Prov}'s functionality embedded provenance graphs. We focus on OCR-APT for THEIA and MAGIC for ATLAS, as these benefit most from functionality features (Table \ref{tab:detection_theia} and \ref{tab:detection_atlas}).

\noindent \textit{\textbf{Finding 12.} MPNet provides the most robust embeddings across heterogeneous logs.}

\noindent MPNet consistently achieves the highest AUC-ROC across attacks and datasets. While DistilBERT often yields comparable results on individual attacks, its performance is less consistent across datasets, whereas RoBERTa is least effective.

\noindent \textbf{LLMs for Functionality Node Features.} We compare different open-source LLMs to generate the functionality features in the 
\textit{node enricher}. We use the LLMs used in the previous investigations, and the prompt is kept constant across all LLMs.\looseness-1 

\noindent \textit{\textbf{Finding 13.} LLaMA-3 produces the most effective functionality features.}

\noindent Fig. \ref{subfig:llm_nodeenricher_comparison} shows that functionality features generated by LLaMA-3 consistently achieve the highest AUC-ROC across most attacks. While other LLMs yield comparable AUC-ROC, this metric alone does not fully capture differences in the quality and coverage of inferred functionality across entities.

\noindent \textit{\textbf{Finding 14.} LLaMA-3 provides the most reliable functionality inference across heterogeneous logs.}


\noindent LLaMA-3 infers functionality for more entities than Qwen-2 and DeepSeek-R1. Fig. \ref{subfig:unclassifiedEntity_comparison} shows that LLaMA-3 fails to infer functionality for only 11 entities, compared to 687 for Qwen-2 and over 3,800 for DeepSeek-R1.
This broader coverage leads to more complete functionality-embedded provenance graphs, which in turn result in more stable detection performance.

\noindent \textit{\textbf{Finding 15.} Behavioral classifier successfully estimates the functionality features for entities unknown to the LLM.}


\noindent 
Although Qwen-2 and DeepSeek-R1 fail to infer functionality for many entities, detectors trained on their outputs achieve AUC-ROC comparable to LLaMA-3 (Fig. \ref{subfig:llm_nodeenricher_comparison}). This shows the effectiveness of the \textit{behavioral classifier} in estimating functionality for unknown entities (Section \ref{subsec:llm_features}). Performance with Qwen-2 is closer to LLaMA-3 due to its smaller set of unknown entities, 
while DeepSeek-R1 relies more on functionality estimation, resulting in lower overall AUC-ROC.





\subsection{Assistant Evaluation} 
\label{subsec:assistant_evaluation}


We evaluate the \textit{graph assistant} on both THEIA \cite{darpa_tc_e3} and ATLAS \cite{alsaheel2021atlas}. For THEIA, we use OCR-APT \cite{aly2025ocr} as the base anomaly detector, as it achieves the largest performance gains with \textit{Auto-Prov} (section \ref{subsec:detection_acc}). For ATLAS, we use MAGIC \cite{jia2024magic} for the same reason. Detected anomalous nodes are linked into graphs using the attack graph construction method from OCR-APT \cite{aly2025ocr}. This choice does not affect the assistant evaluation, as the assistant operates on the resulting graphs rather than on the detection process itself.
The attack graph construction proceeds by selecting the top $n_{seed}$ anomalous nodes (ranked by anomaly score) as seeds and expanding each via a one-hop bidirectional traversal, connecting anomalous nodes through intermediate benign nodes to preserve contextual relationships.

\noindent \textbf{Evaluation.}
In \textit{Auto-Prov}, 
the assistant operates as a post-detection interpretation layer to support and direct the analyst's investigation of potential attacker intent.
Consequently, evaluating the assistant using detection-centric metrics (e.g., recall) would conflate explanation quality with detection performance and fail to capture the assistant’s intended role. Therefore, we evaluate the graph assistant in two ways.

\textit{\textbf{(1) Correctness of tactic reasoning.}}
We evaluate whether the assistant’s \textit{reasoning} for assigning adversary tactics is consistent with established characteristics of those tactics.
For each attack graph, the assistant produces a natural-language summary containing a set of APT tactics, labeled by the assistant, each accompanied by a natural-language \textit{reasoning}.
We compare the assistant’s reasoning against ground-truth descriptions of tactics in the MITRE ATT\&CK \cite{mitre_attack}. 

As this evaluation involves open-ended natural-language reasoning, we adopt \textit{LLM-as-judge} evaluation, following recent work that uses off-the-shelf LLMs to assess reasoning quality when human evaluation is costly \cite{laskar-etal-2025-improving}.
We use three LLMs\textemdash DeepSeek-R1-32B \cite{guo2025deepseek}, Qwen-2-72B \cite{team2024qwen2}, and Gemma-2-27B \cite{team2024gemma}\textemdash as judges, distinct from the assistant’s base LLM to avoid bias. 
Each judge receives (i) the MITRE tactic description and (ii) the assistant’s reasoning, 
and is instructed to determine whether the reasoning reflects the defining characteristics of the tactic.

Formally, let $M$ be the set of tactics labeled by the assistant for a given attack graph. 
For each $m \in M$, let $J_i(m) \in \{0, 1\}$ be the binary judgment of judge $i$, where $1$ indicates correct reasoning and $0$ incorrect. A tactic label is considered correct if at least two out of three judges agree that the reasoning is consistent with the MITRE ground truth. 
We compute an indicator function $\mathbb{I}(m)=1(\Sigma_{i=1}^3J_i(m)\ge 2)$ and define \textit{tactic correctness} 
$\alpha_{TC}=\frac{1}{|M|}\Sigma_{m\in M}\mathbb{I}(m)$. The prompts of the judges are in appendix \ref{subsec:llm_judge_prompts}.

\textit{\textbf{(2) Robustness to unknown entities.}}
To evaluate the assistant's robustness to unseen entities, we apply controlled entity-name poisoning to attack graphs, since the base LLM (LLaMA-3) recognizes most entity names in these popular datasets, hindering systematic evaluation.

For each attack graph, we collect the set of unique entity names $\mathcal{N}_a=\{e_i, \ldots, e_{|\mathcal{N}_a|}\}$. Given a poisoning rate $r\in [0, 100]$, we sample $\mathcal{N}_a' \subseteq \mathcal{N}_a$ with $|\mathcal{N}_a'| = \text{max}(1, \lfloor \frac{r\mathcal{N}_a}{100} \rfloor)$ and define a deterministic mapping $\pi:\mathcal{N}_a' \mapsto \tilde{\mathcal{N}_a}$, where each poisoned name $\tilde{e} = \pi(e)$ is a randomly generated string that preserves the syntactic structure of $e$ (e.g., file extensions) while removing semantic cues. As explained in section \ref{subsec:attack_graph_assistant}, the assistant will flag these poisoned entities as unknown and estimate their functional label using the \textit{behavioral classifier}.

We evaluate robustness along three dimensions. \textit{First}, we measure whether tactic reasoning remains correct under poisoning by computing $\alpha_{TC}$. 
\textit{Second}, we assess explanation stability by comparing summaries generated from the original and poisoned graphs using BERTScore $\alpha_{BERT}$ \cite{zhang2019bertscore, pape2025prompt}, which computes cosine similarity between contextual embeddings from BERT \cite{koroteev2021bert}. \textit{Third}, we evaluate \textit{tactic consistency} by measuring the fraction of tactics in the original summary $M^{o}$ that also appear in the poisoned summary $M^{p}$, $\alpha_{R} = \frac{|M^{o} \cap M^{p}|}{|M^{o}|}$. 
Together, these metrics assess whether the assistant can both reason correctly about adversary tactics and remain robust with unknown entities. We report our findings as follows.


\begin{figure}
\centering
\includegraphics[width=6cm]{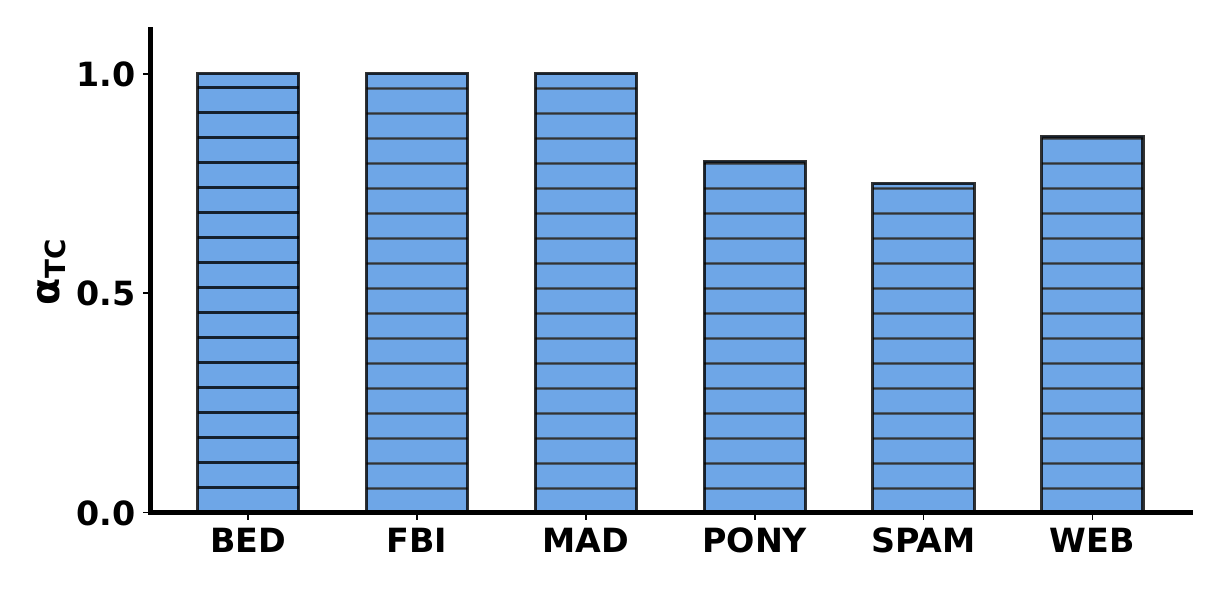}
\caption{\textit{Tactic correctness} across attacks in THEIA \cite{darpa_tc_e3} and ATLAS \cite{alsaheel2021atlas}.}
\label{fig:graph_assist_eval}
\end{figure}

\noindent \textit{\textbf{Finding 16.} The assistant produces reliable tactic interpretations across different attacks.}


\noindent Across all attacks, the assistant achieves consistently high \textit{tactic correctness}, frequently attaining $\alpha_{TC}=1.0$ and remaining above $0.8$ in most cases (Fig. \ref{fig:graph_assist_eval}). 
This indicates that the assistant’s reasoning is largely consistent with the defining characteristics of adversary tactics in the MITRE ATT\&CK. Errors primarily arise in ambiguous system behaviors that admit multiple plausible interpretations. For example, the assistant occasionally maps repeated DNS activity with the ``Initial Access'' tactic, although such patterns may also reflect benign name resolution behavior. These cases lead to lower tactic correctness scores ($e.g., \alpha_{TC} > 0.75$ for SPAM). Importantly, these failures do not dominate overall performance.

\noindent \textit{\textbf{Finding 17.} The assistant remains robust to unknown entities.}



\noindent As shown in \ref{subfig:poison}, under entity-name poisoning, the assistant maintains high \textit{tactic correctness} ($\alpha_{TC}$) and summary similarity ($\alpha_{BERT}$), both consistently exceeding $0.8$ across poisoning rates. This demonstrates that estimated functionality allows the assistant to reason effectively when entities are unknown, producing similar summaries and correct tactic inferences. We observe occasional reductions in \textit{tactic consistency} ($\alpha_R \approx 0.7$), which occur when multiple MITRE tactics exhibit overlapping characteristics rather than due to reasoning errors. Despite this ambiguity, the generated reasoning remains informative and can guide the analyst's investigation.

\section{Discussion and Limitations}
\label{subsec:discussion}


\noindent \textbf{Unseen functionality.} 
\textit{Auto-Prov} infers functionality labels for previously unseen entities using an LLM's knowledge and a similarity-based behavior classifier. This inference is constrained to LLM’s prior knowledge or interaction patterns observed in the log. While entities with entirely new functionality will rarely appear in practice, we note that if a system introduces a novel entity whose functionality is neither recognizable by the LLM nor behaviorally similar to known entities, \textit{Auto-Prov} may fail to assign a correct label. Handling such cases can be a promising direction for future work.


\noindent \textbf{Prompt injection attacks.} 
Auto-Prov assumes that the log inputs and prompts are not adversarially manipulated. An attacker who injects malicious instructions into logs to influence LLM behavior~\cite{chen2025struq} could potentially interfere with provenance extraction or explanation generation. Defending against such attacks is outside of our scope. Existing prompt channeling techniques \cite{chen2025struq} could be incorporated to mitigate this risk.\looseness-1

\section{Conclusion}
\label{subsec:conclusion}

\textit{Auto-Prov} addresses the key limitations in provenance graph based anomaly detection by automating provenance graph construction, embedding functional context for system entities, and providing analyst-oriented attack summaries. 
Across diverse logs, platforms, and state-of-the-art detectors, \textit{Auto-Prov} consistently improves detection accuracy and attack-level coverage while remaining robust to previously unseen entities, advancing the practicality of provenance graph-based anomaly detection in real-world deployments.

\section*{Open Science Policy}
\label{subsec:open_source}

\noindent \textbf{Datasets.} Our experiments are conducted on publicly available datasets~\cite{darpa_tc_e3, alsaheel2021atlas, purseclab_atlas}, 
which have been widely used in prior research. Ground-truth labels are also publicly available~\cite{alsaheel2021atlas,bilot2025sometimes}.

\noindent \textbf{Software Artifacts.} We release our source code at https://github.com/intelligenceafa-cloud/Auto-Prov.
The repository includes scripts and configuration details necessary to reproduce the experiments. 
Additional details on experimental setup are provided in section \ref{sec:evaluation}.

\section*{Ethics Considerations}
\label{subsec:ethics}

We assessed the potential risks and benefits of our research
following the ethical guidelines outlined in \cite{kohno2023ethical}. To the best
of our knowledge, this work adheres to established ethical
standards. Our experiments do not involve sensitive, private,
or personally identifiable information. All experiments are
conducted exclusively on publicly available datasets that were
collected and released in an ethical way for research purposes.

\bibliographystyle{plain}
\bibliography{custom}

\appendix

\section{Appendix}


\subsection{LLM Prompts}
\subsubsection{Candidate Provenance Extractor}
\label{subsec:cpe_prompts}

\noindent
\begin{minipage}{\linewidth}
\begin{findingbox}
    You are an expert in cybersecurity and system provenance graphs. You will be provided with a \textbf{<Platform Name>} log. Your task is to
understand the structure and semantics of the log and produce a structured
natural-language paragraph suitable for constructing a provenance graph.\\

Identify and describe all entities and interactions in the logs that are
relevant to system and network provenance. This may include system components, operating systems, and storage artifacts,
configurations, hosts, users, network endpoints, and, when
present, web-related objects and requests. 
Each entity must be explicitly
named and accompanied by its unique identifier in round brackets.
If identifiers are not provided in the log, assign deterministic sequential
identifiers (id-1) and use them consistently.\\

Summarize only provenance interactions (e.g., events, accesses,
requests) between entities, and exclude irrelevant identifiers or low-level
log artifacts that are not useful for provenance graph construction.\\

Instructions:\\

1. Read the log carefully and identify all interactions relevant to provenance graph construction. Write a single paragraph describing all activities in the log, where each sentence corresponds to one interaction.
   
2. If present, include only entity names enclosed in quotes "..." with
   their identifiers in round brackets (...).
   
3. Mark interaction types using curly brackets \{...\}.

4. Mark entity types (if present) using square brackets [...].

5. Preserve subject and object identifiers explicitly and avoid introducing
   any identifiers beyond those required for provenance construction.
   
6. Strictly follow the output format below.\\

Output Format:

\textbf{<single structured paragraph>}
\end{findingbox}
\vspace{-0.5em}
\centering \textbf{Prompt 1:} Log Summarization
\end{minipage}

\noindent
\begin{minipage}{\linewidth}
\begin{findingbox}
You are an expert in cybersecurity and system provenance graphs. You will be given as input (i) a \textbf{<Platform Name>} log and (ii) a 
[SUMMARY] describing the activities observed in that log.\\

Your task is to extract all identifiers relevant to provenance graph
construction and the corresponding entity types. This
includes identifiers explicitly mentioned in the paragraph as well as any
additional identifiers present in the log that are necessary to represent
provenance-relevant entities or network endpoints.\\

First, identify all unique entity identifiers appearing in round brackets (...),
ignoring event-specific or transient identifiers. In addition, identify any
network addresses and ports appearing in the log or paragraph and distinguish
their roles when possible (e.g., local vs. remote, source vs. destination).\\

Then, for each extracted identifier, determine the system-level type of the corresponding
entity by using the entity-type annotations in the paragraph or
implicit type information available in the log. If no provenance-relevant
entity type can be determined, assign the type NONE. Multiple identifiers may
share the same entity type.\\

Strictly follow the output format below.\\

Output Format:\\

"ID-1" = "entity-type-1"

"ID-2" = "entity-type-2"

"ID-3" = "NONE"

"IP:port-1" = "source address"

"IP:port-2" = "destination address"
\end{findingbox}
\vspace{-0.5em}
\centering \textbf{Prompt 2:} Entity Types Extraction
\end{minipage}

\noindent
\begin{minipage}{\linewidth}
\begin{findingbox}
You are an expert in cybersecurity and system provenance graphs. 
You will be given as input:
(1) A \textbf{<Platform Name>} log  
(2) A [SUMMARY] describing the activities observed in that log\\

Your task is to extract all meaningful relationships and resolve entity names for 
provenance graph construction, following two consecutive steps:\\

Step A: Identify Relationships Between Entities and IPs\\

- Read the log and [SUMMARY] to identify all unique entity identifiers in round brackets (...) 
  and all network addresses and ports. 
  
- Detect transitory relationships (e.g., parent-child between services or files) 
  and explicit interactions between entities.
  
- Identify network connections between IP addresses and between entities and IPs.

- For each detected relationship, generate a unique pair in one of the formats: (ID, ID), (IP:port, IP:port), 
  or (ID, IP:port).
  
- If no meaningful relationships exist, output: [NO MEANINGFUL PAIRS POSSIBLE]\\

Step B: Resolve Entity Names\\

- For each entity ID appearing in the relationship pairs, locate the corresponding full entity name 
  (e.g., system-level service, application, files, URL) in the paragraph.
  
- If no full name is available,
  assign NONE.
  
- Ignore network addresses; focus only on system entities.\\

Output Format:\\

[RELATED ENTITIES and IP ADDRESSES]\\

(ID-1, ID-2)  A: [...]

(ID-1, ID-3)  A: [...]

(ID-2, ID-5)  A: [...]

(IP:port-1, IP:port-2)  A: [...]

(ID-6, IP:port-5)  A: [...]\\

[ENTITY NAMES]\\

"ID-1" = "entity-name-1"

"ID-2" = "entity-name-2"

"ID-3" = NONE

"ID-4" = "entity-name-4"

"ID-5" = NONE

\end{findingbox}
\vspace{-0.5em}
\centering \textbf{Prompt 3:} Entity Extraction
\end{minipage}

\noindent
\begin{minipage}{\linewidth}
\begin{findingbox}
You are an expert in cybersecurity and system provenance graphs. You will be given as input:

1. A \textbf{<Platform Name>} log  

2. A [SUMMARY] describing the activities observed in the log  

3. Pairs of entity IDs as (ID, ID)

4. Pairs of IP addresses as (IP:port, IP:port)  

5. Pairs of entity IDs and IPs as (ID, IP:port) or (IP:port, ID)  \\

Your task is to determine the action/event/access/interaction types or web requests occurring between these entities, along with their directions and timestamps, to construct a provenance graph. Perform this step by step:\\

Step 1: Analyze the log and [SUMMARY] line by line to understand all operations. 

Step 2: For each pair, fill the "A: [...]" with the action/event/access types or web requests as mentioned in the log. Do not rephrase, invent, or omit any relevant types. If no type is found, write "NO LABEL". 

Step 3: Determine the direction of the interaction: \{D=->\} for left-to-right, \{D=<-\} for right-to-left.  

Step 4: Include the timestamp of each interaction as (timestamp=...). If the log does not provide a timestamp, leave it as (timestamp=...).  

Step 5: Strictly follow the output format below.  \\

Output Format:\\

(ID-1, ID-2)  A: [...] \{D=-> or D=<-\} (timestamp=...)  

(ID-1, ID-3)  A: [...] \{D=-> or D=<-\} (timestamp=...)  

(ID-2, ID-5)  A: [...] \{D=-> or D=<-\} (timestamp=...)  

(ID-6, ID-3)  A: [...] \{D=-> or D=<-\} (timestamp=...)  

(IP:port-1, IP:port-2)  A: [...] \{D=-> or D=<-\} (timestamp=...)  

(ID-6, IP:port-5)  A: [...] \{D=-> or D=<-\} (timestamp=...)  

\end{findingbox}
\vspace{-0.5em}
\centering \textbf{Prompt 4:} Edge Extraction
\end{minipage}

\subsubsection{Rule Generator}
\label{subsec:rule_generator_prompt}

\noindent
\begin{minipage}{\linewidth}
\begin{findingbox}

You are an expert in system provenance and regular expressions. You are given:
(1) a raw system log entry, and
(2) a single field value from a provenance record extracted from that log
(e.g., source ID, destination ID, interaction type, entity name, or timestamp).\\

Your goal is to induce a regex rule that, when applied to
the raw log text, reproduces the same field value. Rules must be structurally anchored, reusable across logs with the same format, and extract only the intended value. Reason step by step, but your output must strictly follow the required format.\\

\textbf{TASK 1.}
Given an exact identifier value (e.g., source ID, destination ID, entity ID) or
Vtype, generate a \emph{generic and
reusable} regex that extracts this same value from the raw log.
If the given value consists of multiple components (e.g., IP address and port),
generate separate regex patterns for each component. \\

\textbf{TASK 2.}
Given an extracted interaction type value, generate a \emph{generic and
reusable} regex that extracts any interaction type appearing at the same structural location in similar logs.
Do not include the specific extracted value in the regex; match only the format

\textbf{TASK 3.}
Given an extracted timestamp, generate a \emph{generic and
reusable} regex that extracts timestamps
from the same structural location in similar logs.
Do not include the specific timestamp value in the regex; match only the format.\\

\textbf{TASK 4.}
Given an extracted entity name (e.g., domain, URL, file path, or filename),
generate a \emph{generic and
reusable} regex that extracts any entity name from the same structural
location in similar logs.\\

\textbf{Rule Validation and Correction Guidelines.}
All generated regex rules must extract \emph{only} the intended value and nothing
else. Preserve the punctuation, spacing, and
capitalization. 
Use capturing groups \texttt{()} only around the intended value.
Never use overly broad patterns (e.g., \texttt{.+}) to capture content.
If a generated regex is found to be incorrect, fix it by preserving the exact log
structure and generalizing only the specified token using \texttt{[A-Za-z]+}.

If a task cannot be completed unambiguously from the log, return \texttt{No Regex}.

\end{findingbox}
\vspace{-0.5em}
\centering \textbf{Prompt 5:} Rule Generator
\end{minipage}

\subsubsection{Node Enricher}
\label{subsec:node_enricher_prompt}

\noindent
\begin{minipage}{\linewidth}
\begin{findingbox}

You are an expert in cybersecurity and system provenance graphs. You will be given a text from \textbf{<Platform Name>} logs that may represent entities such as files, directory, URL/website, or command-line invocation, etc. Your task is to classify the entity, analyze its purpose and functionality, and provide a detailed type label. Do this step by step.\\

Step 1: Determine the broad category of the entity: If it is a file (has extensions like .exe, .dll etc.), a directory, classify as a **File/Directory**. If it is a URL (e.g., http://, www.), classify as **URL/Website**. If it is a command-line invocation, classify as  **Command-Line**. If it is abstract, ambiguous, classify it as **NO LABEL**.\\

Step 2: Examine the entity in detail: For **File/Directory**: analyze the extension, path structure, naming patterns, and likely system usage (library, config, system file, etc.). For **URL/Website**: analyze the protocol, domain, subdomain; consider what service it represents. For **Command-Line**: analyze its arguments and parameters; determine the command’s purpose and effect in the system.\\

Step 3: Contextual Analysis: Consider the entity’s role in system operations. What would the file, directory, URL, or command typically do? How does it interact with the system or network?\\

Step 4: Functional Explanation: Write a detailed explanation of what the entity does, its purpose, and how it operates. Explain each technical term if applicable.\\

Step 5: Type Label: Summarize the functionality into a concise but detailed type label. If the entity is invalid or ambiguous, assign `NO LABEL`.\\
Strictly follow the output format below:\\

\textbf{entity\_name | Type: <functional\_label>}

\textbf{entity\_name | Type: <NO LABEL>}

\end{findingbox}
\vspace{-0.5em}
\centering \textbf{Prompt 6:} Assigning Functional Labels $\tau_f$
\end{minipage}

\subsubsection{Graph Assistant}
\label{subsec:graph_assistant_prompts}

\noindent
\begin{minipage}{\linewidth}
\begin{findingbox}

You have been given a system entity (eg., processes, files, network sockets). Based on your knowledge, do you have information about the functionality of this system entity?

Respond with ONLY "YES" or "NO" nothing else.

\end{findingbox}
\vspace{-0.5em}
\centering \textbf{Prompt 7:} Flagging unknown nodes
\end{minipage}

\noindent
\begin{minipage}{\linewidth}
\begin{findingbox}

You are an expert in cybersecurity and system provenance graphs. You are given
a graph in which each line represents a directed edge of the form:\\

    [source node] --[interaction type]--> [destination node]\\

Edges are ordered chronologically from earliest to latest, reflecting the
temporal sequence of events. When the same edge (same source node, destination
node, and interaction type) appears consecutively, it is annotated with a
repetition count, e.g., (x3), indicating repeated operations such as repeated
connections, file access, or requests.
You are given: A list of nodes that have been detected as malicious. Supplementary node metadata in the form of (node name, node functionality), which describes the system-level functionality of some nodes appearing in the
  graph.The graph may represent only a portion of a larger attack.\\

\textbf{Task 1: Provenance-Based Narrative Generation}
Produce a detailed natural-language narrative that describes what is happening
in this graph step by step, based solely on the observed entities, interactions,
and their temporal order. Construct the most coherent explanation possible from
the available evidence, without inferring external context or attack phases.

Return one paragraph in the following format:

    Summary: \textbf{<single continuous narrative paragraph>}\\

Guidelines for Task 1: Do not include headings, labels, bullet points, or numbered lists inside the summary text. Do not speculate beyond what is supported by the graph structure and node metadata.\\

\textbf{Task 2: APT Tactic Identification}
You are given a list of APT tactics.
Based only on the events described in the generated summary, identify which tactics are explicitly matches the observed behavior. List only
the tactics that are evident from the summary; do not infer tactics that are not in the list.
For each inferred tactic, provide a concise explanation that explicitly
references the system entities and interactions described in the summary.

Use the following structured format:

    Stage: \textbf{<APT tactic name>}
    
    Reasoning: \textbf{<evidence-based explanation>}\\

Guidelines for Task 2: You do not need to identify all tactics, only those clearly supported by the
  summary. Base your reasoning strictly on the narrative produced in Task 1. Avoid speculation or assumptions beyond the described evidence.

\end{findingbox}
\vspace{-0.5em}
\centering \textbf{Prompt 8:} Summarizing attack graphs and labeling APT tactics. 
\end{minipage}

\subsubsection{LLM Judges}
\label{subsec:llm_judge_prompts}

\noindent
\begin{minipage}{\linewidth}
\begin{findingbox}

You are an expert and evaluator in cybersecurity and the MITRE ATT\&CK framework. Below, you are given two pieces of reasoning: (1) Model Reasoning, and (2) MITRE Reference. Carefully compare the Model Reasoning to the MITRE Reference Reasoning.\\

    \textbf{Task:}
    Does the Model Reasoning align with the MITRE Reference for the model to categorize this attack under this APT tactic?\\
    
    Format your answer as:
    YES/NO

\end{findingbox}
\vspace{-0.5em}
\centering \textbf{Prompt 9:} LLM Judge Prompt
\end{minipage}

\begin{figure}[t]
\centering
\includegraphics[width=\linewidth]{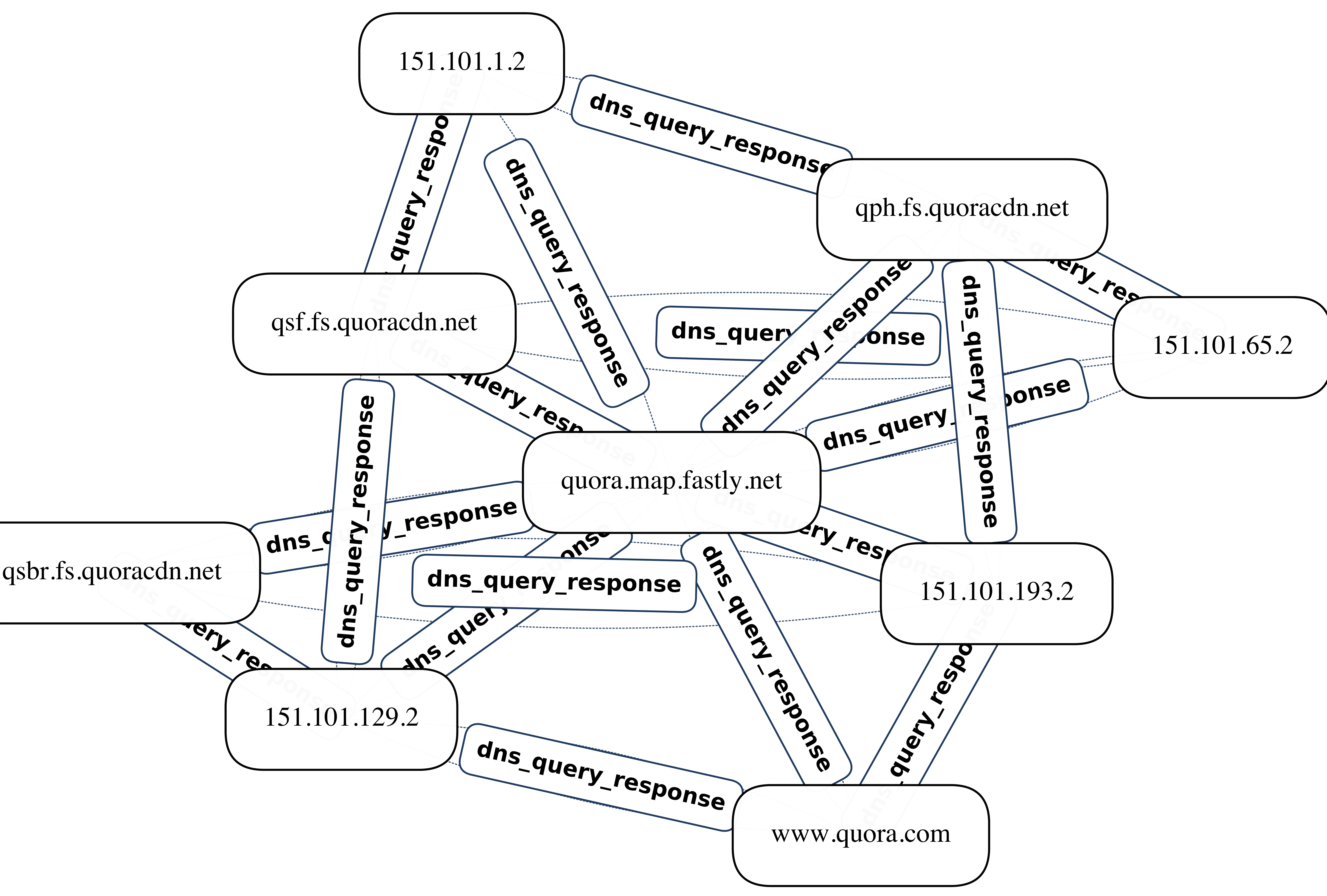}
\caption{An example of \textit{malvertising dominate} attack graph from ATLAS \cite{alsaheel2021atlas}.}
\label{fig:attack_graph_example}
\end{figure}

\subsection{Graph Assistant Summaries}
\label{subsec:graph_assistant_outputs}

In Fig. \ref{fig:graph_assistant_output}, the generated narrative closely mirrors the detected attack graph in Fig. \ref{fig:attack_graph_example}, explicitly describing repeated DNS queries and responses while preserving their temporal ordering. The underlying graph exhibits high edge density, frequent repeated interactions, and many-to-many communication patterns across entities. Instead of simply restating these patterns as suspicious, the graph assistant contextualizes them as consistent with DNS resolution and CDN caching behavior.

This behavior is valuable from an analyst’s perspective. While anomaly detectors are sensitive to irregular patterns, its difficult to know why an alert. The graph assistant offers plausible benign interpretations of the observed interactions, enabling analysts to quickly assess whether an alert warrants deeper investigation or can be deprioritized. 

The assistant’s utility is further reflected in its mapping to APT tactics. Based on the summarization, it identifies only \textit{Reconnaissance} and explicitly rules out subsequent tactics. The accompanying reasoning is evidence-based, emphasizing the absence of actions beyond DNS queries. This restrained tactic attribution helps analysts avoid over-interpreting early-stage or ambiguous interactions. 



\begin{figure}[t]
\centering
\includegraphics[width=\linewidth]{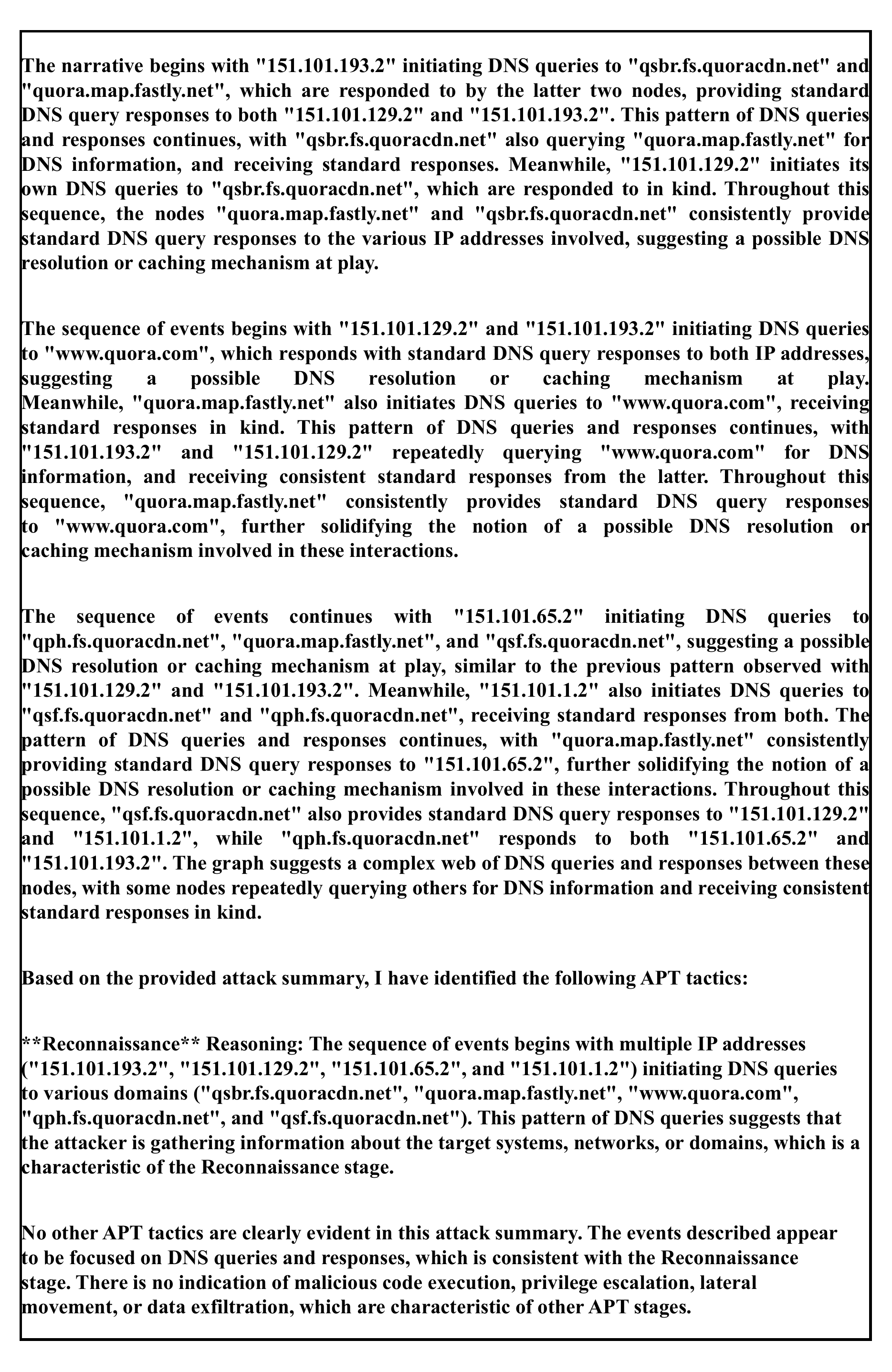}
\caption{An example of graph assistant's output on the attack graph in Fig. \ref{fig:attack_graph_example}.}
\label{fig:graph_assistant_output}
\end{figure}





\end{document}